\documentclass[
a4paper,
preprint,
amsmath,
aps,
pra, showpacs,
]{revtex4-1}

\usepackage[english]{babel}
\usepackage[utf8]{inputenc}
\usepackage{graphicx}
\usepackage[colorinlistoftodos]{todonotes}

\date{\today}

\begin{document}

\title{Can one turn off Coulomb focusing?}

\date{\today}

\author{S.A.\ Berman}
\affiliation{School of Physics, Georgia Institute of Technology, Atlanta, Georgia 30332-0430, USA}
\affiliation{Centre de Physique Th\'eorique, CNRS--Aix-Marseille Universit\'e, Campus de Luminy, 13009 Marseille, France}

\author{C.\ Chandre}
\affiliation{Centre de Physique Th\'eorique, CNRS--Aix-Marseille Universit\'e, Campus de Luminy, 13009
Marseille, France}

\author{T.\ Uzer}
\affiliation{School of Physics, Georgia Institute of Technology, Atlanta, Georgia 30332-0430, USA}

\begin{abstract}
We find that Coulomb focusing persists even when the Coulomb field is barely noticeable compared with the laser field. Delayed recollisions proliferate in this regime and bring back energy slightly above the $3.17~U_p$ high-harmonic cutoff, in stark contradiction with the Strong Field Approximation. We investigate the nonlinear-dynamical phase space structures which underlie this dynamics. It is found that the energetic delayed recollisions are organized by a reduced number of periodic orbits and their invariant manifolds.
\end{abstract}


\maketitle

\section{Introduction}

The theoretical framework for electron dynamics processes in intense linearly polarized laser fields was established about twenty years ago and remains the state-of-the-art in strong field physics~\cite{Kuch87,Cork93,Scha93}. It centers on the ``recollision'' model which follows a three-step scenario: Electrons are first ionized, absorb energy while following the laser, and then are propelled back to the ionic core after the laser reverses direction, about half a laser cycle after ionization. The kinetic energy transfer in the core region causes ionization of more electrons (nonsequential multiple ionizations - NSMI)~\cite{Beck12} or generates very high harmonics of the driving laser by high harmonic generation (HHG)~\cite{Lewe94}. Despite the wide-spread use of the three-step model, it is not widely appreciated that it usually takes more than one recollision for the ionized electron to transfer its energy to set the processes of multiple ionization or high-harmonic generation into motion. It is reasonable to expect that the spreading of the electronic wave packet would make these delayed collisions much less effective. Surprisingly, the opposite is the case, and it is due to a process called Coulomb focusing~\cite{Brab96}. The  effectiveness of Coulomb focusing was demonstrated convincingly in the pioneering work of Brabec {\em et al.} \cite{Brab96}, who showed that the nonsequential double ionization rate is enhanced by an order of magnitude by Coulomb focusing.

Research in the last two decades has confirmed that Coulomb focusing (and the Coulomb field in general) is a key player in the recollision process~\cite{Shaf12,Shaf13,Comt05,Boha02,Sand99}.
But it seems that the opposite question --namely, when the Coulomb focusing does not matter-- has not received the same attention. Perhaps the reason is that the answer seems obvious: the Coulomb field can be ignored when the laser field is strong. After all, this is one of the tenets of the Strong Field Approximation (SFA)~\cite{Ivan05} which ignores the Coulomb field when the electron is moving in the laser field. But how strong is ``strong'' for the laser field, and how weak does the Coulomb field have to be for us to neglect it altogether --i.e., is it possible to turn it off completely, as the three-step model would have us do?

In fact, the Coulomb field and its focusing effect persist well beyond what would be expected based on the comparison of field strengths, and well beyond field strengths where there is no potential barrier to trigger tunneling ionization at peak field. In this range of intensity, contrary to SFA predictions, delayed recollisions continue to manifest themselves and bring back energy slightly above the $3.17~U_p$ high-harmonic cutoff. The kinetic energy brought back by delayed recollisions is compatible with the value provided in the SFA for immediate recollision (occurring half a laser cycle after ionization), but those energetic delayed recollisions do not exist in the SFA. 

Our paper is structured as follows: In Sec.~\ref{sec2} we begin by reviewing how the return energy is maximized in the SFA, and characterize the trajectories bringing maximum energy to the core. 
In Sec.~\ref{sec3} we investigate how the energy-maximizing classical recollision trajectories change when the Coulomb field is included. We show that delayed energy-maximizing trajectories, which are not present in the SFA, emerge. We provide a characterization of the delayed recollisions in phase space, and show that they are related to the invariant manifolds of particular periodic orbits.

\section{Maximizing the Return Energy in the Strong Field Approximation}
\label{sec2}

	We begin with the derivation of some results on recollisions in the Strong Field Approximation (SFA). These results will form the basis for understanding the distinct effect of the Coulomb field on recollision trajectories.
The SFA may be described by the one-dimensional classical Hamiltonian (in atomic units)
\begin{equation}\label{hammie}
H(x,p,t) = \frac{1}{2}\Big(p + \frac{E_0}{\omega}\cos(\omega t + \phi)\Big)^2\, ,
\end{equation}
where $p$ is the canonical momentum, $E_0$ is the electric field amplitude, $\omega$ is the laser frequency, and the dipole approximation for the laser field is used.
In the context of the three-step model, this Hamiltonian describes the motion of the electron after its ionization, which we define as occurring at $t=0$ without loss of generality (up to a change of phase $\phi$).
The three-step model does not specify the exact phase of the laser at the instant of ionization, so we allow this phase $\phi$ to be a free parameter of the model.

	We write Hamiltonian \eqref{hammie} in the velocity gauge as opposed to the more common length gauge to make the translational invariance of the SFA apparent (i.e., independence of Hamiltonian \eqref{hammie} with respect to $x$).
Thus, it is clear that the momentum $p$ is conserved, and as a consequence, the resulting system is integrable. It is convenient to rescale the variables so that $x$ is in units of the quiver radius $E_0/\omega^2$, $p$ is in units of $E_0/\omega$, and $t$ is in radians, so Hamiltonian \eqref{hammie} now looks like 
\begin{equation} \label{hammie scaled}
\mathcal{H}(x,p,t) = \frac{1}{2}\Big(p + \cos(t + \phi)\Big)^2.
\end{equation}
We also note that this Hamiltonian has the symmetry
\begin{equation}\label{symmetry}
(x,p,\phi) \rightarrow (-x,-p,\pi+\phi),
\end{equation}
which allows us to narrow our focus to initial conditions with $\phi \in [0,\pi)$.

	From Eq.\ \eqref{hammie scaled} we write down the solution for $x(t)$,
\begin{equation} \label{x(t)}
x(t) = x_0 - \sin \phi + p_0t + \sin(t+\phi),
\end{equation}
and $p(t)=p_0$.
Just as the three-step model does not exactly specify the laser phase $\phi$ at ionization, it also does not specify the position of the electron $x_0$ at ionization.
Thus, we allow $x_0$ to be a free parameter as well.
Also, we note that Eq.\ \eqref{x(t)} shows that $p_0$ is the constant drift momentum of the electron.

	Equation \eqref{x(t)} is used to compute the maximum return energy of the electron.
First, it is assumed that the electron reaches the atomic core at time $t_r$, i.e.\ $x(t_r) = 0$.
By substituting this into Eq.\ \eqref{x(t)}, one obtains \begin{equation}\label{p0}
p_0(t_r,\phi) = \frac{\sin\phi - \sin(t_r+\phi) - x_0}{t_r}.
\end{equation}
Now, one may fix $x_0$ and maximize the energy at return, \begin{equation}\label{return energy}
\varepsilon(t_r,\phi) = \frac{1}{2}\Big(p_0 + \cos(t_r+\phi)\Big)^2,
\end{equation}
with respect to $t_r$ and $\phi$.
Setting the derivatives equal to zero gives
\begin{align} \label{dtr}
\partial_{t_r}\varepsilon & = \Big(p_0+\cos(t_r+\phi)\Big)\Big(\partial_{t_r}p_0-\sin(t_r+\phi)\Big) = 0, \\ \label{dphi}
\partial_{\phi}\varepsilon & = \Big(p_0+\cos(t_r+\phi)\Big)\Big(\partial_{\phi}p_0-\sin(t_r+\phi)\Big) = 0.
\end{align}
We discard the solution $p_0+\cos(t_r+\phi)=0$ which corresponds to the zero-energy minima of Eq.\ \eqref{return energy}. Thus, we get $\partial_{t_r}p_0 = \partial_\phi p_0$, and carrying out these derivatives using Eq.\ \eqref{p0} immediately gives
\begin{equation}\label{p0eq0}
\dot{x}(t = 0) = p_0 + \cos\phi = 0,
\end{equation}
which corresponds to zero initial momentum in the length gauge. Though this result has been known for $x_0=0$, we have shown that it is independent of this condition and is valid for any $x_0 \neq 0$.

	In addition, Eq.\ \eqref{p0eq0} tells us that when the return energy is maximized, the initial energy is zero and thus minimized, implying that the change in the energy
$$\Delta \varepsilon = \frac{1}{2}\bigg[\Big(p_0+\cos(t_r+\phi)\Big)^2-\Big(p_0+\cos\phi\Big)^2\bigg],
$$
is at a maximum as well.
By rewriting $\Delta \varepsilon$ as 
\begin{equation}\label{dE calc}
\Delta \varepsilon = \Big(p_0+\frac{1}{2}\cos(t_r+\phi) + \frac{1}{2}\cos\phi\Big)\Big(\cos(t_r+\phi)-\cos\phi\Big),
\end{equation}
substituting Eq.\ \eqref{p0} into Eq.\ \eqref{dE calc}, and using some trigonometric identities, it is shown that
\begin{align}\label{dE}
\Delta \varepsilon (t_r,\phi) & = f(t_r)\sin(t_r+2\phi) + g(t_r,\phi) x_0, \\ \label{f(t_r)}
\text{where } f(t_r) & = \frac{\sin\frac{t_r}{2} \Big(2\sin\frac{t_r}{2} - t_r\cos\frac{t_r}{2}\Big)}{t_r}, \\ \label{g}
\text{and } g(t_r,\phi) & = \frac{2 \sin \frac{t_r}{2} \sin ( \frac{t_r}{2} + \phi)}{t_r}.
\end{align}
We first treat the case of the electron starting very close to the core, i.e.\ $x_0=0$.
In that case Eq.\ \eqref{dE} simplifies to
\begin{equation}
\label{dE0}
\Delta \varepsilon_0(t_r,\phi) = f(t_r)\sin(t_r+2\phi).
\end{equation}
To compute the maximum of this function, we first note that all the $\phi$ dependence is contained in the multiplicative sine term.
Also, one may find a $\phi$ such that $\sin(t_r+2\phi)=\pm1$ for any $t_r$.
Therefore, the absolute maximum of $\Delta \varepsilon_0$ occurs where $|f(t_r)|$ is at its absolute maximum for $t_r > 0$, and $\sin(t_r+2\phi)$ is the sign of $f(t_r)$ at the extremum.
Then we obtain the absolute extremum of $f(t_r)$ in Eq.\ \eqref{f(t_r)} by setting $\partial_{t_r}f = 0$ and using trigonometric double angle identities, yielding
\begin{equation}\label{extremum}
2 - 2t_r\sin t_r+(t_r^2-2)\cos t_r = 0,
\end{equation}
in agreement with the literature~\cite{Cork93,Lewe94,Band05}.
This equation's first positive root, $t_{r,0}^* \approx 4.09$, is the absolute extremum of $f(t_r)$ for $t_r > 0$, and $f(t_r) > 0$.
Thus, we obtain the exact relation between the return time and the initial phase $\phi_0^*$ maximizing $\Delta \varepsilon_0$,
\begin{align} \label{phase relation}
& \sin(t_{r,0}^*+2\phi_0^*)  = 1, \\ \nonumber
\text{or equivalently, } & t_{r,0}^* + 2\phi_0^* = \frac{\pi}{2} + 2\pi n,\,\text{for } n\geq 0.
\end{align}
For $n=1$, this gives $\phi_0^* \approx 1.88$.
Substituting $(t_{r,0}^*,\phi_0^*)$ into Eq. \eqref{dE0}, we finally get the maximum change in energy $\Delta \varepsilon_0^* \approx 3.17~U_p$, where $U_p = E_0^2/4\omega^2$ or simply $1/4$ in rescaled units.
	
    Remarkably, we have found a very simple relation in Eq.\ \eqref{phase relation} between the initial phase of the laser and the return time by maximizing the change in energy instead of the return energy itself.
This relation is borne out by the numerical results for the local maxima of $\dot{x}(t_r,\phi)$ shown in Table 1 of Ref.~\cite{Band05}, when taking care to subtract $\pi/2$ from each of the phases in that table to reconcile our choice of $\sin$ versus $\cos$ for the laser field term in Hamiltonian~\eqref{hammie scaled}.
All odd multiples of $\pi/2$ appear in that table because local extrema are being considered, and $f(t_r) < 0$ for the minima, requiring $\sin(t_r+2\phi) = -1$.
Equation \eqref{phase relation} follows naturally from maximizing $\Delta \varepsilon$, and in our view, $\Delta \varepsilon$ is the fundamental quantity to maximize because it is the energy gained by the electron from the laser field.
Thus, we will only consider maximizing $\Delta \varepsilon$ for the remainder of this paper.

	Now we will consider the effect of allowing $x_0 \neq 0$ on the trajectories that maximize $\Delta \varepsilon$ in the SFA, because in the three-step model the electron's tunnel ionization implies its starting position is not exactly $x_0 = 0$.
Though the consequences of $x_0 \neq 0$ are investigated in Ref. \cite{Band05}, these simulations include the Coulomb field.
Therefore it is not clear whether the observed effects are due to the inclusion of the Coulomb field or the relaxation of the $x_0 = 0$ condition, and we intend to clarify this.
We have already shown that $\dot{x}(t=0) = 0$ will persist when $x_0 \neq0$, but to see the effect on the other quantities $\Delta \varepsilon$ in Eq.\ \eqref{dE} must be maximized.
If we restrict $x_0 \ll 1$, then we can treat the effect of $x_0 \neq 0$ on the maximization of $\Delta \varepsilon$ perturbatively.
This is not an unreasonable restriction because $x_0$ just needs to be small compared to the quiver radius.
For example, if one assumes that ionization occurs around $1$ a.u., the laser intensity is $I=6 \times 10^{13}\, \text{W}\cdot\text{cm}^{-2}$, and the laser wavelength is $800$ nm, the corresponding value of $x_0$ in rescaled units is $x_0 \approx 0.08$.
For the same frequency but an intensity of $I = 10^{16}\,\text{W}\cdot\text{cm}^{-2}$, which is used for numerical simulations in this paper, the corresponding value of $x_0$ in rescaled units is even smaller: $x_0 \approx 6 \times 10^{-3}$.
\begin{figure}
\centering
\hspace*{-0.6in}
\includegraphics[height=0.37\textheight]{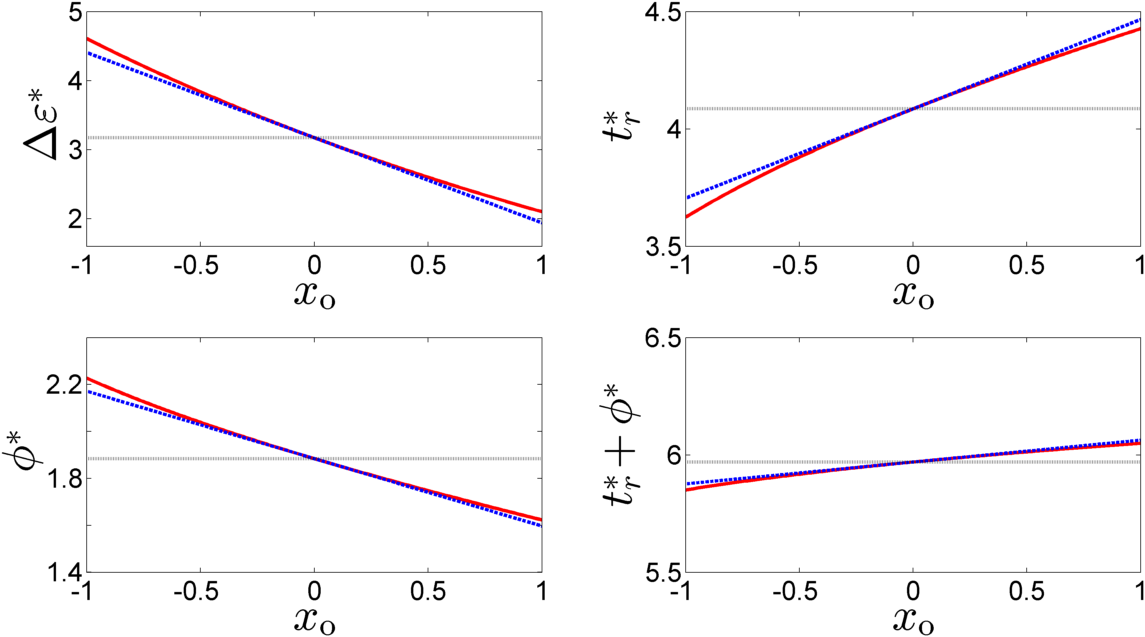}
\caption{(color online) Dependence on $x_0$ of the maximum energy exchange $\Delta \varepsilon^*$ (upper left panel), return time $t_r^*$ (upper right panel), laser phase $\phi^*$ (lower left panel) and $t_r^*+\phi^*$ (lower right panel) from Hamiltonian \eqref{hammie scaled}. Solid (red) curves are the numerical values and the dashed (blue) lines are the first order approximations. Grey dotted lines are the values for $x_0 = 0$. $x_0$ is in units of $E_0/\omega^2$, $\Delta \varepsilon^*$ is in units of $U_p$, $t_r^*$ and $\phi^*$ are in radians.}
\label{x0neq0}
\end{figure}
To do the calculation, we assume first order perturbation expansions of $t_r^*$ and $\phi^*$ in integral powers of $x_0$,
\begin{align} \label{pert t}
t_r^* & = t_{r,0}^* + t_1 x_0 + \mathcal{O}(x_0^2), \\ \label{pert phi}
\phi^* & = \phi_0^* + \phi_1 x_0 + \mathcal{O}(x_0^2),
\end{align}
with the zeroth order terms taking their values from the $x_0=0$ case.
We carry out the calculation to first order by computing $(t_1,\phi_1)$.
Recalling Eqs.\ \eqref{g} and \eqref{dE0} and setting the partial derivatives of Eq.~\eqref{dE} with respect to $t_r$ and $\phi$ equal to $0$, we get
\begin{align}\label{dtr2}
\partial_{t_r}(\Delta \varepsilon_0) + x_0\, \partial_{t_r}g  & = 0, \\ \label{dphi2}
\partial_{\phi}(\Delta \varepsilon_0) + x_0\, \partial_{\phi}g & = 0.
\end{align}
Now we substitute the perturbation expansions \eqref{pert t} and \eqref{pert phi} into Eqs.\ \eqref{dtr2} and \eqref{dphi2} and discard terms of order $\mathcal{O}(x_0^2)$.
By Taylor-expanding the derivatives of $\Delta \varepsilon_0$ to first order in $x_0$ and applying phase relation \eqref{phase relation} to the $(t_{r,0}^*,\phi_0^*)$ terms, we get the system of linear equations
\begin{equation}\label{t1phi1}
\begin{pmatrix}
-\partial^2_{t_rt_r}(\Delta \varepsilon_0)\big|_0 & 2\Delta \varepsilon_0^* \\
2\Delta \varepsilon_0^* & 4\Delta \varepsilon_0^*
\end{pmatrix}
\begin{pmatrix}
t_1\\
\phi_1
\end{pmatrix} = 
\begin{pmatrix}
\partial_{t_r}g \\
\partial_{\phi}g
\end{pmatrix}\bigg|_0,
\end{equation}
where ``$|_0$'' means to evaluate at the parameters from the $x_0=0$ case. Equation~\eqref{t1phi1} yields $(t_1,\phi_1) \approx (0.38,-0.29)$.

	The results of these approximations are compared against numerical maximization of Eq.~\eqref{dE} in Fig.~\ref{x0neq0}.
These approximations do work well for $|x_0| \ll 1$, and even appear to do a fair job closer to $|x_0| = 1$.
Note that the shift of the final phase $t_r+\phi$ is slightly less pronounced than for $t_r$ and $\phi$, as its first order coefficient is small, $t_1+\phi_1 = 0.09$, so that $t_r^*+\phi^*$ appears fairly constant.

We notice on the upper left panel of Fig.~\ref{x0neq0} that $\Delta \varepsilon^*$ significantly exceeds the well-known value of $3.17~U_p$. This has already been noted in Ref.~\cite{Band05} [and is also easily seen from Eq.~\eqref{dE calc} using Eq.~\eqref{p0eq0}]: The maximum is expected to be at $8~U_p$ which occurs at $x_0=\pi$ (e.g., with parameters $\phi=\pi$ and $t_r=3\pi$ with $p_0=1$). However only the region close to the core is physically relevant since the electron is initially bound to the core. Therefore we do not expect significant variations of initial laser phase, ionization time and maximum return energy from the values obtained from $x_0=0$. In summary, the typical energy-maximizing recollision occurs within one laser cycle and brings an amount of energy close to $3~U_p$. 
When the laser field is large, we expect a similar result by taking into account the Coulomb field, and this is what we are going to investigate in the next section.

% \subsection{Phase Space Perspective of SFA}
%
\begin{figure}
\includegraphics[width=0.45\textwidth]{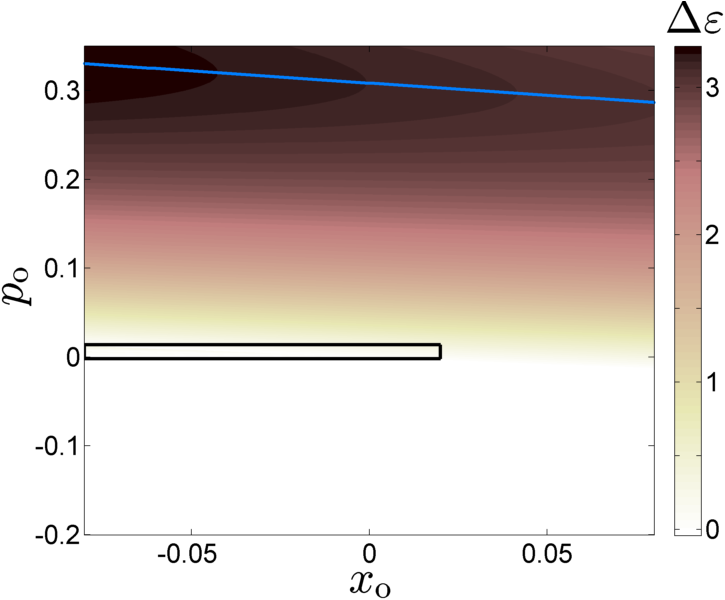}
\includegraphics[width=0.45\textwidth]{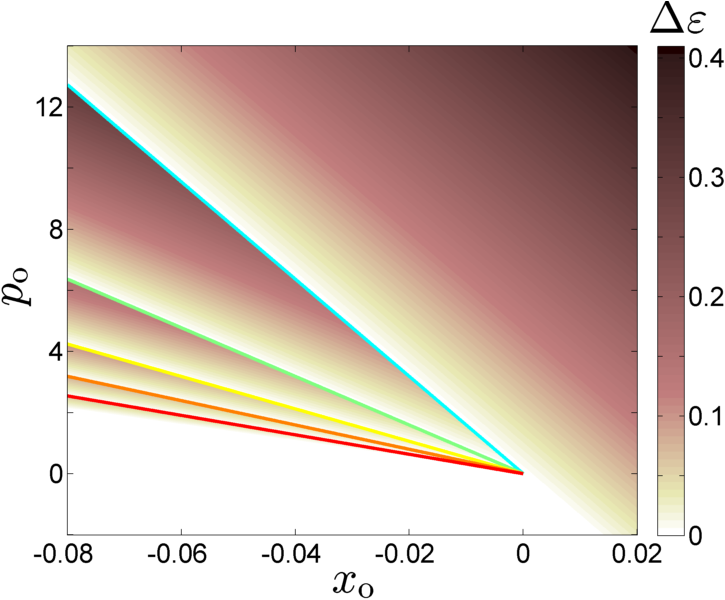}
\caption{(color online) Energy exchange $\Delta \varepsilon$ at first recollision (after a six laser cycle integration) as a function of initial condition, $(x_0,p_0)$. The initial phase $\phi$ is fixed by Poincar\'e section condition \eqref{poincare condition}. Darker points correspond to higher energy. Left panel: The initial conditions $(x_0,p_0)$ bringing the maximum energy $\Delta \varepsilon$ at first recollision for each $x_0$ are the light grey (blue) curve. Right panel: Zoom of the black-framed region on the left. Solid grey (colored) lines separate regions of initial conditions that have their first recollision during different laser cycles, with the colors corresponding to Fig.~\ref{fig:sfaPS}. $x_0$ is in units of $E_0/\omega^2$, $p_0$ is in units of $E_0/\omega$, $\Delta \varepsilon$ is in units of $U_p$.}\label{poincare energies SFA}
\end{figure}

When examining the effect of the Coulomb field on the energy-maximizing trajectories, we need to study the location of their initial conditions in phase space in detail.
Before proceeding it is enlightening to look at the phase space picture of energy-maximizing trajectories in the SFA.
% Were these delayed recollisions present in the SFA picture? Actually there are delayed recollision in the SFA picture.
In order to locate these trajectories in phase space, we consider a Poincar\'e section of the trajectories. %in SFA, i.e., obtained from Hamiltonian \eqref{hammie scaled}. 
In the left panel of Fig.~\ref{poincare energies SFA}, we consider a grid of initial conditions $(x_0,p_0)$ beginning on the Poincar\'{e} section
\begin{equation}\label{poincare condition}
\dot{x} = 0 \,\, \text{with} \,\, \ddot{x} < 0,
\end{equation}
and integrated for six laser cycles, with $x_0 \in [-0.08, 0.08] E_0/\omega^2$ and $p_0 \in [-0.2,0.35] E_0/\omega$.
% The trajectories are given by Eq. \eqref{x(t)}.
We notice that the Poincar\'e section determines the initial phase such that $\cos \phi=-p_0$ and $\sin \phi > 0$.
Since $\dot{x}=p_0+\cos(t+\phi)$, the Poincar\'e section is a stroboscopic plot with period $2\pi$ in the SFA. 
The condition $\dot{x}=0$ is a natural section because we have just shown that all of the trajectories maximizing $\Delta \varepsilon$ satisfy this condition in the SFA.
For all trajectories beginning on this Poincar\'{e} section, the electron begins at rest and starts to move to the left.
The reversal of the laser field causes the electron to reverse its direction and move to the right, towards the core again.
It is then possible for a recollision to occur, and thus we define a ``recollision" as crossing $x = 0$ moving to the right, i.e.\ with $\dot{x} > 0$.

In Fig.~\ref{poincare energies SFA}, we show the energy $\Delta \varepsilon$ brought back by each initial condition on its first recollision.
When starting at the top of the left panel of Fig.~\ref{poincare energies SFA} and moving down, we observe a continuous gradient of $\Delta \varepsilon$, until a series of discontinuous changes beginning at small $p_0 > 0$ and $x_0 < 0$.
These discontinuities are seen clearly in the right panel of Fig.~\ref{poincare energies SFA}, which shows that the first discontinuous change in $\Delta \varepsilon$ is a sudden increase in energy, followed by another continuous decreasing gradient of $\Delta \varepsilon$.
This behavior repeats over and over as $p_0$ gets closer and closer to $0$.
It turns out that these sudden changes in energy as the initial conditions are varied correspond to changes to the laser cycle during which an initial condition has its first recollision, referred to as delayed recollisions in what follows as opposed to immediate recollisions which occur during the first laser cycle. The delayed recollisions are analogous to the ``higher-order returns'' of Ref.~\cite{Brab96}.
%On the contrary, when we turn on the Coulomb field we will show below that these trajectories can also carry high energy, actually exceeding the $3.17 U_p$ prediction of the SFA.
% 
\begin{figure}
\setlength{\unitlength}{1cm}
\begin{picture}(11,7)
\put(0,0){\includegraphics[width=0.6\textwidth]{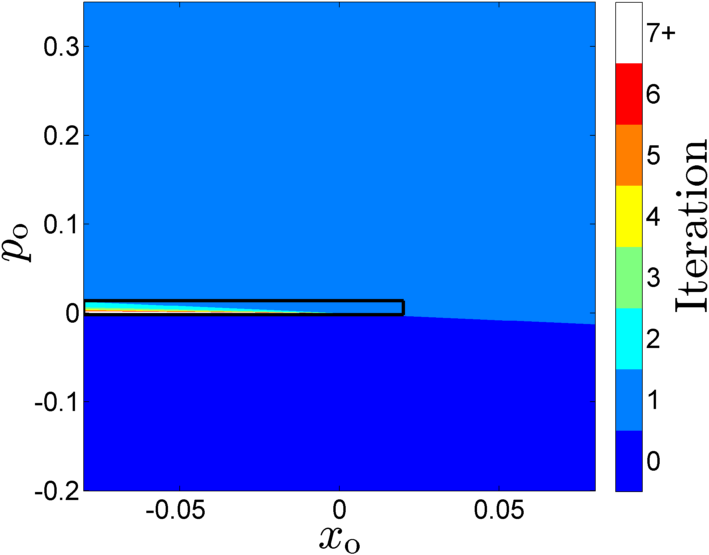}}
\put(4.1,4.25){\includegraphics[width=0.25\textwidth]{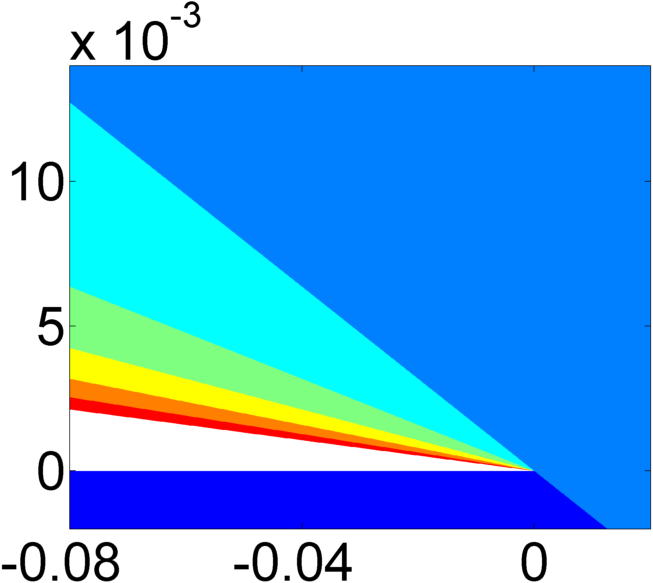}}
\end{picture}
\caption{\label{fig:sfaPS} (color online) Number of iterations of Poincar\'e map $\mathcal{P}$ until the first recollision has occurred, as a function of initial conditions $(x_0,p_0)$, where $\phi$ is fixed by the Poincar\'{e} section condition \eqref{poincare condition}. In the SFA, a single iteration of this map always corresponds exactly to the duration of one laser cycle. $x_0$ is in units of $E_0/\omega^2$, $p_0$ is in units of $E_0/\omega$.}
\end{figure}
In Fig.~\ref{fig:sfaPS} we determine which laser cycle the first recollision will occur for a given initial condition analytically using Eq.~\eqref{x(t)} and the Poincar\'e section condition \eqref{poincare condition}. The general structure in Fig.~\ref{fig:sfaPS} is that immediate recollisions in the ``1" region are separated from the immediate ionizations in the ``0'' region by a small triangle in which the delayed recollisions live.
A closer look at the delayed recollision triangle in the inset of Fig.~\ref{fig:sfaPS} shows that the triangle is bounded below by the line $p_0=0$ and its rightmost point is $(x_0,p_0)=(0,0)$.
Furthermore, the triangle is clearly stratified, with each stratum corresponding to a delayed recollision with a particular delay.
The strata are well ordered in phase space: as $p_0$ decreases, the delay time increases.
These observations are explained by considering the mechanical scenario in this region:
All trajectories starting on the section will first move to the left, and as is evident from Eq.~\eqref{x(t)}, they will eventually come to rest (at approximately half of the laser cycle for small $p_0$), move to the right, and come to rest again, this time on the Poincar\'e section (with exactly one laser cycle elapsed).
They will have a displacement from their initial position of $\Delta x = 2\pi p_0$.
Therefore, trajectories beginning with $x_0 < 0$ and a small enough $p_0 > 0$ such that $x_0 + \Delta x < 0$ will not experience a recollision in the first laser cycle: they must wait for enough laser cycles $n$ such that the displacements accumulate and satisfy $x_0 + n \Delta x > 0$.
The smaller the $p_0$, the more laser cycles are needed for a recollision to occur, which explains the ordering of the strata.
This argument also explains why there are no recollisions if $x_0 < 0$ and $p_0 < 0$, and none for $x_0 > 0$ and $p_0 < -x_0/2 \pi$.
The lines that form the boundaries between the delayed recollisions of differing delay times are plotted in the right panel of Fig.~\ref{poincare energies SFA}.
We note that the boundaries occur precisely where the return energy has discontinuous jumps, as claimed earlier.

It should be noted that the highest energies associated with delayed recollisions are an order of magnitude smaller than the immediate recollisions, which is reflected in the change of color scale between the left and right panels of Fig.~\ref{poincare energies SFA}.
Thus, when searching for energy-maximizing trajectories in the SFA, these kinds of trajectories appear irrelevant, not bringing enough energy to play a significant role in recollision-driven processes.
However it should be emphasized that this is only on their \emph{first} recollision.
In the SFA, these trajectories with very small $p_0$ will have many recollisions, as is evident from Eq.\ \eqref{x(t)}.  
With $\dot{x} = p_0 + \cos(t + \phi)$, the return kinetic energy of the later recollisions can only get slightly higher than $2~U_p$, still significantly below the usual prediction of $3.17~U_p$.
Moreover, the maximum return kinetic energy of any recollision (i.e.\ not necessarily the trajectory's \emph{first} recollision) occurring after the first laser cycle for a trajectory initiated near $x=0$ is bounded at approximately $2.4~U_p$~\cite{Band05}.
This changes when we turn on the Coulomb field: even if the laser field is large, the Couloumb field allows delayed recollisions to bring back energy greater than $3.17~U_p$ on their very first recollision. In addition, the trajectories experiencing delayed recollision are particularly interesting because they spend a longer time close to the core where they can exchange energy and experience the influence of the core potential.

\section{Maximizing the return energy with the Coulomb field: Nonlinear dynamics of delayed recollisions}
\label{sec3}
\subsection{Delayed Recollisions}
\begin{figure}
\includegraphics[width=0.5\textwidth]{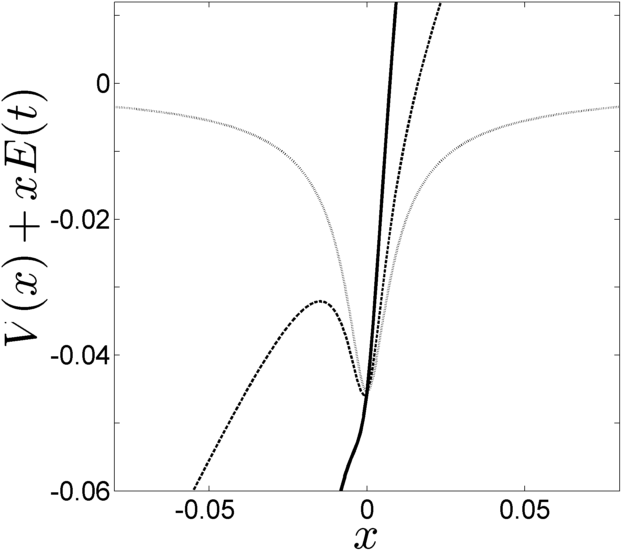}
\caption{The potential of Hamiltonian \eqref{hammie scaled coulomb} in the length gauge, i.e.\ $V(x) + x E(t)$, as a function of $x$. Grey dotted is with $E(t) = 0$, black dashed with $E(t) = 0.25~E_0$ (the maximum field amplitude when $I = 6.3 \times 10^{14} \, \mathrm{W}\cdot\mathrm{cm}^{-2}$), and solid black with $E(t) = E_0$. $x$ is in units $E_0/\omega^2$ and $V$ is in units of $U_p$.}\label{fig:coulomb well}
\end{figure}

    In Sec.~\ref{sec2}, we have identified the effects of $x_0 \neq 0$ on the recollision mechanism in the strong field approximation. In particular we have assessed the effect of $x_0 \neq 0$ on the maximum $\Delta \varepsilon$ trajectories. We have shown that the energy-maximizing trajectories are the ones which recollide within one laser cycle. In this section, we turn on the Coulomb interaction and investigate how the strong field approximation picture for recollision trajectories is affected by this interaction.
The Hamiltonian, in atomic units, now becomes
\begin{equation}\label{hammie coulomb}
H(x,p,t) = \frac{1}{2}\Big(p + \frac{E_0}{\omega}\cos(\omega t + \phi)\Big)^2 - \frac{1}{\sqrt{x^2+1}} \, ,
\end{equation}
where we have used a soft-Coulomb potential with a softening parameter of $1$~\cite{Java88,Beck12}.
The general phase space picture associated with Hamiltonian \eqref{hammie coulomb} is composed of two distinct regions~\cite{Maug09_1}: the region near the core, referred to as the bounded region, where the Coulomb attraction dominates compared to the laser forcing, and the region sufficiently far from the core, referred to as the unbounded region, where the laser field dominates.
For low enough laser intensities, the electron never escapes the bounded region. A common assumption in the unbounded region is to neglect the Coulomb interaction for sufficiently large intensity, which corresponds to the second step in the three-step model of recollisions.
Here we investigate the validity of this assumption, using numerical simulations of the equations of motion arising from Hamiltonian \eqref{hammie coulomb}.
In order to focus on the effect of the Coulomb field in the unbounded region, we choose the parameters in such a way that the bounded region is completely suppressed.
In particular, we use a laser intensity $I = 10^{16}\, \mathrm{W}\cdot\mathrm{cm}^{-2}$ and frequency $\omega \approx 0.057$ a.u. (corresponding to an $800$ nm wavelength).
At this laser frequency, the bounded region only exists for intensities below $I = 4.9 \times 10^{15}\, \text{W}\cdot\text{cm}^2$.
Thus, our choice of laser intensity is high enough that almost every initial condition will lead eventually to ionization, meaning there are almost no trajectories that remain indefinitely bounded to the core.
In fact, this intensity is so high that there is not even a potential barrier when the laser field is at its maximum, as shown in Fig.~\ref{fig:coulomb well}, so classical ionization can account for all ionizations; it is not necessary to invoke a tunneling argument.

	We rescale this Hamiltonian as in Sec.~\ref{sec2}, yielding
\begin{equation} \label{hammie scaled coulomb}
\mathcal{H}(x,p,t) = \frac{1}{2}\Big(p + \cos(t + \phi)\Big)^2 - \frac{\epsilon}{\sqrt{x^2+a^2}},
\end{equation}
where $a = \omega^2/E_0 \approx 6.1 \times 10^{-3}$\ and $\epsilon = \omega^4/E_0^3 \approx 6.9 \times 10^{-5}$. The parameter $\epsilon$ is the effective strength of the Coulomb potential in the presence of the laser field: as the intensity goes to infinity, $\epsilon$ goes to zero as does $\epsilon/a$ so the Coulomb potential becomes negligible.
In this parameter regime it would appear that the Coulomb potential is ignorable due to the small magnitude of the maximum Coulomb energy, $\epsilon/a \approx 0.046~U_p$.
However, the relative importance of terms in the Hamiltonian is not determined by the absolute values of the terms themselves but by the absolute values of the gradients of the terms, because it is the gradients that actually appear in the equations of motion.
In this parameter regime, the Coulomb force, equal to $-\epsilon x/(x^2+a^2)^{3/2}$ has a maximum amplitude of $2\epsilon/(3a^2\sqrt{3})\approx 0.7 E_0$, which is comparable to the maximum electric force, and therefore cannot be neglected.

In the same way as we did in Sec.~\ref{sec2}, we examine the trajectories that maximize the change in energy $\Delta \varepsilon$ between some starting position $x_0$ at $t = 0$ and the final position at the center of the core, $x=0$, at the return time $t_r$.
With the inclusion of the Coulomb field, the translational invariance of the SFA is broken, meaning the momentum $p$ is no longer conserved.
Thus, this system is no longer integrable, so trajectories cannot be obtained analytically. Here we numerically integrate the equations of motion associated with Hamiltonian \eqref{hammie scaled coulomb} in order to find the trajectories that maximize $\Delta \varepsilon$.
We notice that the discrete symmetry \eqref{symmetry} remains, meaning we may still confine our simulations to initial conditions with $\phi \in [0,\pi)$.
Using a symplectic integrator \cite{Ruth83}, we again consider a large number of initial conditions (typically of the order of one thousand for each value of $x_0$) beginning on the Poincar\'e section \eqref{poincare condition}, and look for the trajectories that yield the highest $\Delta \varepsilon$ at the first recollision.
The range of initial conditions is the same as in the previous section, and with this choice of laser parameters the ranges of initial conditions in atomic units correspond to $x_0 \approx [-13,13]$ a.u. and $p_0 \approx [-1.9,3.3]$ a.u.
\begin{figure}
\includegraphics[width=0.92\textwidth]{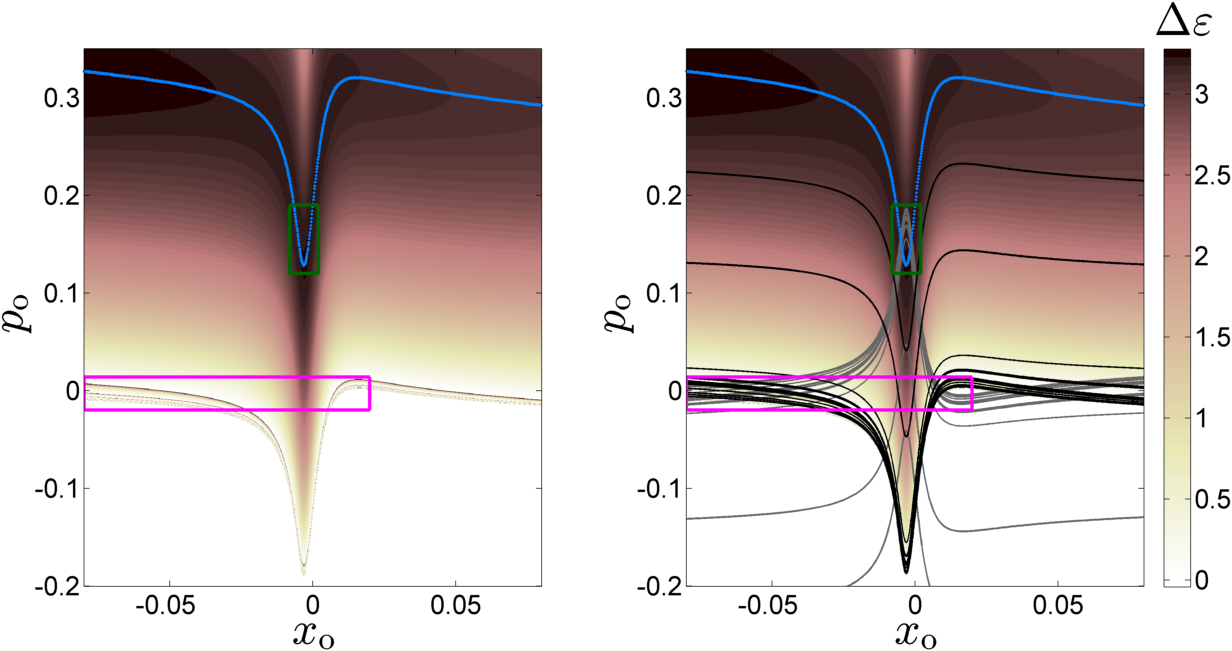}
\includegraphics[width=0.92\textwidth]{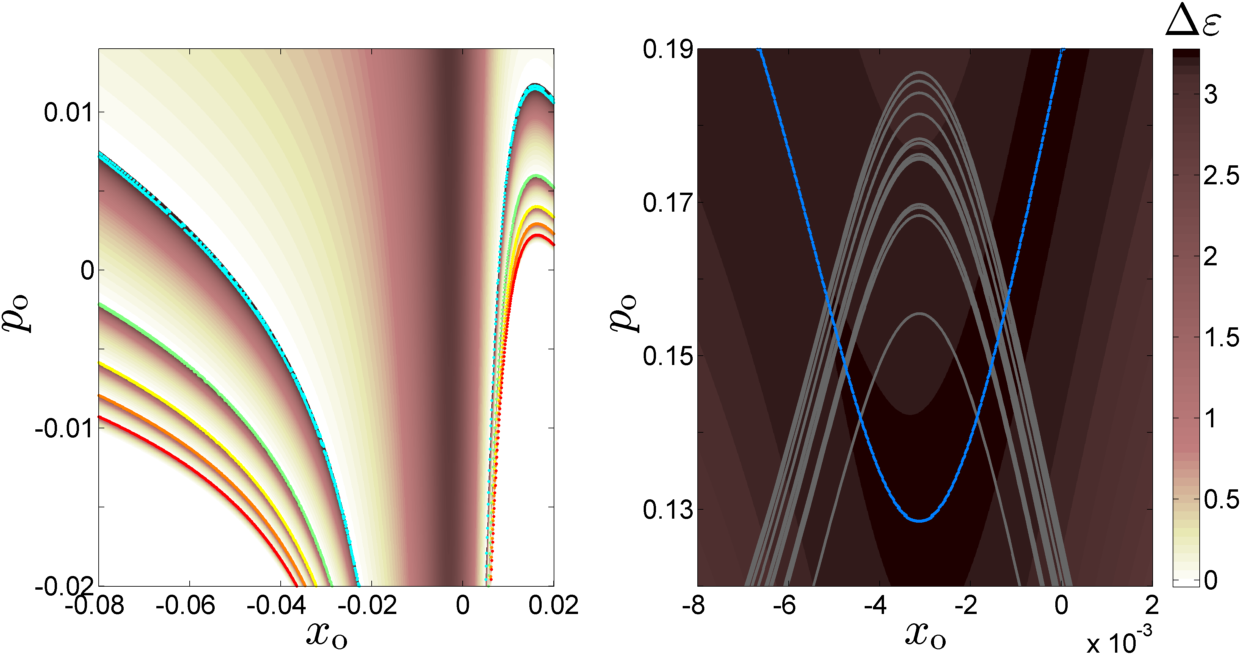}
\caption{(color online) Energy exchange $\Delta \varepsilon$ at first recollision after a six laser cycle integration of Hamiltonian \eqref{hammie scaled coulomb} with $a \approx 6.1 \times 10^{-3}$\ and $\epsilon \approx 6.9 \times 10^{-5}$, as a function of initial conditions, $(x_0, p_0)$ with the initial phase $\phi$ is fixed by Poincar\'e section condition \eqref{poincare condition}. Darker points correspond to higher energy. Top left panel: The initial conditions $(x_0,p_0)$ bringing the maximum energy $\Delta \varepsilon$ at first recollision in the first laser cycle for each $x_0$ are the light grey (blue) curve.
Top right panel: Same as top left, with the stable manifold $\mathcal{W}^s$ of $\mathcal{O}$ plotted in black and its unstable manifold $\mathcal{W}^u$ in grey. Bottom left panel: Zoom of the lower grey (magenta) framed region on top. Grey (colored) curves are the initial conditions $(x_0,p_0)$ bringing the maximum energy at first recollision for each $x_0$ and iteration of Poincar\'e map $\mathcal{P}$ during which the recollision happens. The colors correspond to Figs.~\ref{fig:sfaPS} and \ref{poincare iterations}. Bottom right panel: Zoom of upper grey (green) framed region above. $x_0$ is in units of $E_0/\omega^2$, $p_0$ is in units of $E_0/\omega$, $\Delta \varepsilon$ is in units of $U_p$.}
\label{poincare energies}
\end{figure}

\begin{figure}
\includegraphics[width=0.5\textwidth]{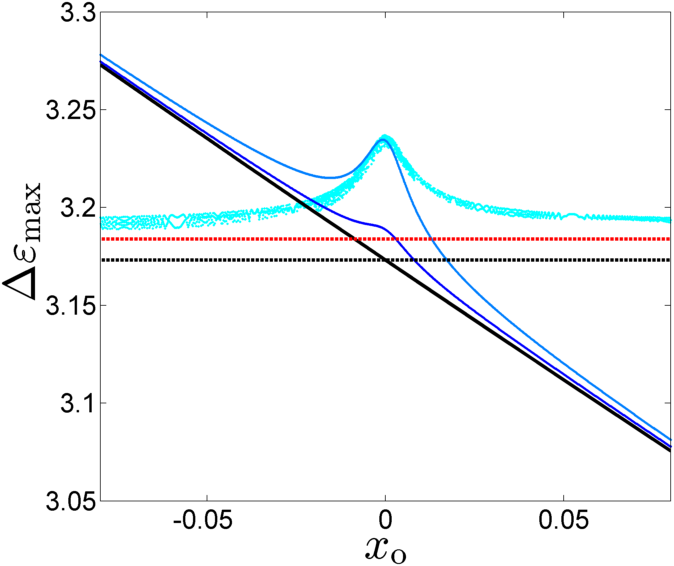}
\caption{(color online) Maximum change in energy $\Delta \varepsilon$ as a function of $x_0$. Thick black line: SFA, immediate recollision. Grey (light blue) curve: Coulomb included, immediate recollision. Light grey (cyan points): Coulomb included, second cycle delayed recollision. Thin black (blue) curve: Coulomb included, immediate recollision final energy, $\varepsilon(t_r)$. Dotted black line: 3.17 $U_p$. Dotted grey (red) line: Estimate of delayed recollision energy ``cutoff.'' $x_0$ is in units of $E_0/\omega^2$, $\varepsilon$ is in units of $U_p$.}\label{fig:dEvsx0}
\end{figure}

In the top left panel of Fig.~\ref{poincare energies}, we show the energy $\Delta \varepsilon$ brought back by each initial condition on its first recollision.
Comparing this with Fig.~\ref{poincare energies SFA}, we can evaluate the effect of the Coulomb field on the recolliding trajectories.
The shape of the curve of initial conditions leading to $\Delta \varepsilon$-maximizing immediate recollisions is deformed compared to the SFA, due to the presence of the Coulomb field.
The actual energies of these recollisions as a function of $x_0$ are seen more clearly in Fig.~\ref{fig:dEvsx0}.
We see that the immediate recollision curve is mostly close to the SFA curve, especially as $|x_0|$ increases.
At such a high laser intensity, its deviations from the SFA curve are in large part due to the leading Coulomb correction to the final return energy $\varepsilon (t_r)$ \cite{Kamo14}, which is indeed even closer to the SFA curve.
Therefore, we now see clearly the differing effects of varying $x_0$ versus turning on the Coulomb field on the return energy.
If one considers the total energy at return (as opposed to the energy difference or only the return kinetic energy), then the effect of the Coulomb field on the value of the return energy compared with the SFA is small for this laser intensity.
The major deviations from $3.17~U_p$ come from varying $x_0$, and in the previous section we account for these effects using the SFA.

\begin{figure}
\includegraphics[width=0.95\textwidth]{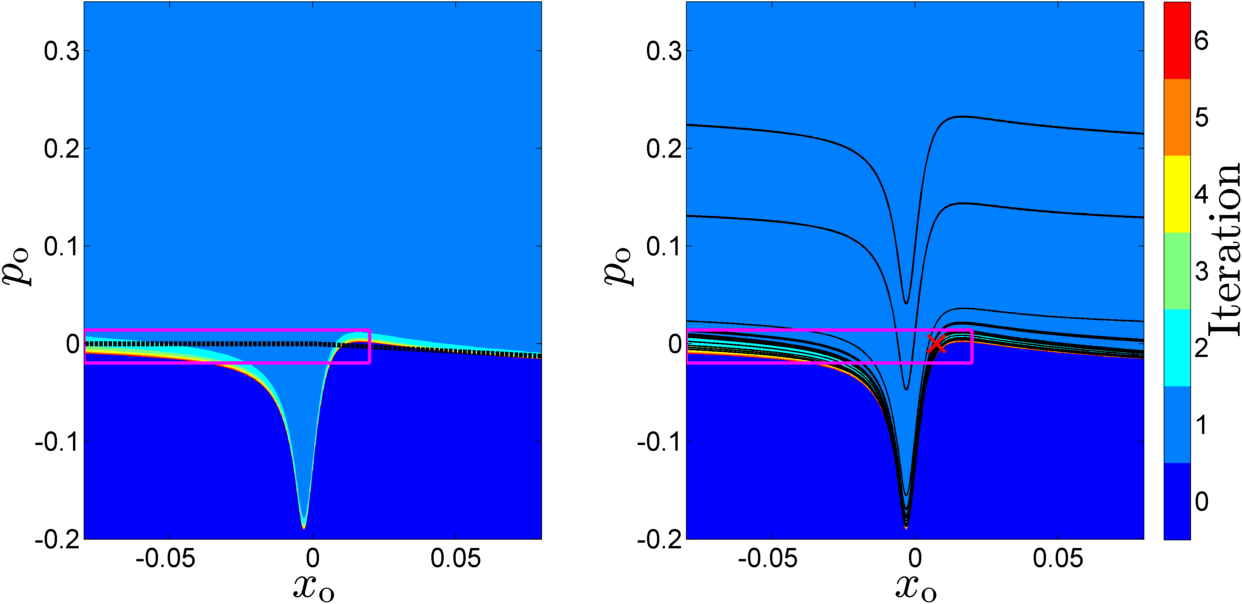}
\includegraphics[width=0.5\textwidth]{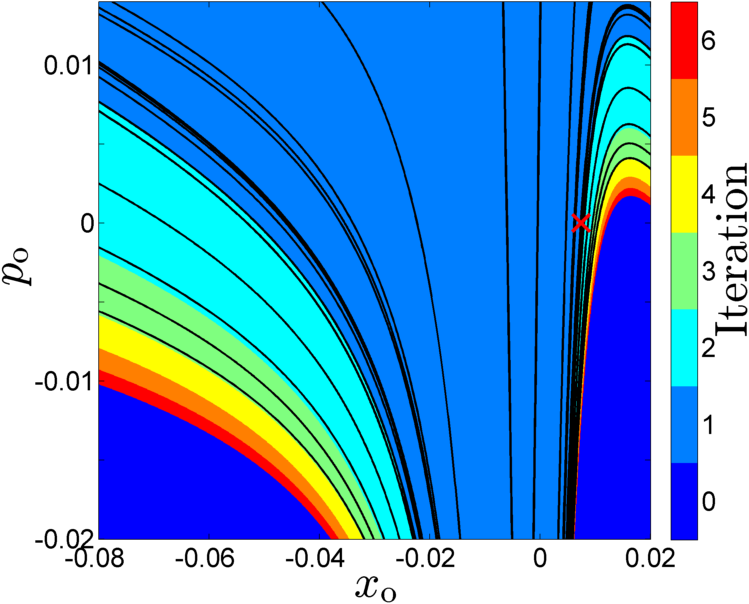}
\caption{(color online) Number of iterations of Poincar\'e map $\mathcal{P}$ until the first recollision has occured after a six laser cycle integration, as a function of initial condition $(x_0,p_0)$. The initial phase $\phi$ is fixed by Poincar\'e section condition \eqref{poincare condition}. Top left panel: The black dotted line separates immediately ionizing trajectories from recolliding trajectories in the SFA. Top right panel: Same as top left panel, with stable manifold $\mathcal{W}^s$ of $\mathcal{O}$ plotted in black. Location of $\mathcal{O}_-$ marked as the grey (red) cross. Bottom: Zoom of the grey (magenta) framed region above. $x_0$ is in units of $E_0/\omega^2$, $p_0$ is in units of $E_0/\omega$.}
\label{poincare iterations}
\end{figure}

Now we reexamine the delayed recollisions with the Coulomb field on.
The continuous gradient of $\Delta \varepsilon$ in the upper left panel of Fig.~\ref{poincare energies} near the maximum energy immediate recollisions is again interrupted, at low values of $p_0$, by a series of jumps, seen more closely in the bottom left panel of Fig.~\ref{poincare energies}.
This is where the delayed recollisions originate, which can be confirmed by inspecting Fig.~\ref{poincare iterations}.
Like Fig.~\ref{fig:sfaPS}, Fig.~\ref{poincare iterations} shows the iteration of the Poincar\'e map during which each initial condition has its first recollision.
The Poincar\'e map is the discrete map $(x_{i+1},p_{i+1}) = \mathcal{P}(x_i,p_i)$ that takes a point $(x_i,p_i)$ on the Poincar\'e section \eqref{poincare condition} to the next point $(x_{i+1},p_{i+1})$ at which the trajectory started with $(x_i,p_i)$ pierces the surface of section.
Thus, for the range of initial conditions we consider, the iteration of the map $\mathcal{P}$ during which an initial condition $(x_0,p_0)$ has its first recollision is the smallest $n$ such that $x_n > 0$.
As we stated earlier, in the SFA the map $\mathcal{P}$ corresponds exactly to the duration of one laser cycle.
With the Coulomb field on and a large laser intensity, this correspondence continues to hold only approximately.
Still, for the remainder of the paper, when we refer to a ``cycle'' during which a recollision occurs, we really mean the iteration $n$ of the Poincar\'e map $\mathcal{P}$.
Comparing Fig.~\ref{poincare iterations} and Fig.~\ref{fig:sfaPS} makes the significant qualitative differences between the SFA and the full Hamiltonian \eqref{hammie scaled coulomb} apparent.
For example, the Coulomb focusing effect is clearly manifested by the greater area of phase space that leads to recollisions compared with the SFA picture.
The curve that separates immediate ionizations from recolliding trajectories in Fig.~\ref{fig:sfaPS} is plotted on the upper panel of Fig.~\ref{poincare iterations}.
We see that a substantial area of the immediate ionizations in the SFA are converted into recolliding trajectories with the Coulomb field on, especially in the core region (small $x_0$).
In particular, the delayed recollisions are enhanced by the Coulomb field.
In the previous section we observe from Fig.~\ref{fig:sfaPS} that the delayed recollision initial conditions lay between the immediate recollisions and immediate ionizations in the region of phase space with a small $p_0 > 0$ and $x_0 < 0$.
With the Coulomb field on, we see in Fig.~\ref{poincare iterations} that the delayed recollision boundary extends to $x_0 > 0$ and near $x_0=0$ it takes a significant dip in the direction $p_0 <0$, so the presence of the Coulomb field enables more delayed recollisions.

In addition, the Coulomb field allows certain delayed recollisions to bring energy that slightly exceeds the well known $3.17~U_p$  upon their very first recollision.
This is shown in the lower left panel of Fig.~\ref{poincare energies}, where although we have changed the distribution of colors on the color scale compared with the panel above it, the overall scale remains the same.
It is true that most of each recollision region of a particular delay contains trajectories that first recollide with a low amount of energy close to that brought by delayed recollisions in the SFA, $\lesssim 0.5~U_p$.
However, the energy at first recollision rapidly increases near the boundaries of the recollision regions of differing delays.
The initial conditions bringing back the maximum energy for each $x_0$ and delay time are the colored curves.
Actually these curves are made up of discrete points (one for each $x_0$ on our grid) that in most places are so close together that they appear to form a continuous curve.
The energies of the second-cycle delayed recollisions as a function of $x_0$ are plotted in Fig.~\ref{fig:dEvsx0}.
We observe that for any $x_0$, there exists a second-cycle delayed recollision that brings back energy exceeding $3.17~U_p$.
Moreover, the amount of energy brought back by these delayed recollisions does not display the approximately linear dependence of $\Delta \varepsilon_\text{max}$ on $x_0$ characteristic of the immediate recollisions, but rather approaches a constant value for large $|x_0|$ and has a jump near $x=0$.
Surprisingly, the Coulomb well is all but invisible at maximum laser amplitude, yet its presence causes certain delayed recollisions to recollide with an energy above the usual SFA prediction on their very first recollision.
Though we do not show the maximum energies brought by higher order delayed recollisions in Fig.~\ref{fig:dEvsx0}, we argue in the next section that for every $x_0$ in the range we consider, there is a recollision of arbitrary delay that carries the same energy as we observe the maximum energy second-cycle recollisions carry~\footnote{We omitted the maximum energies of the higher order delayed recollisions because the shrinking size of the regions of phase space leading to recollisions with increasingly longer delays (see Fig.~\ref{poincare iterations}) makes it challenging to accurately compute the initial condition that recollides with the maximum energy for a particular delay.}.
Therefore, during every laser cycle there are energetic recollisions with energies close to $3.17~U_p$, and this is due to the Coulomb field.
We emphasize that the SFA misses this completely - as discussed in Section~\ref{sec2}, the energetic immediate recollisions are ionized after one laser cycle, and only trajectories with sufficiently low drift momenta will continue to recollide with energies $\lesssim 2.4~U_p$.
In the next section, we explain the nonlinear dynamical origin of the delayed recollisions, and the origin of the cutoff in their energy exchange.

\subsection{Building Blocks of Recollisions: Periodic Orbits and Their Invariant Manifolds}
\begin{figure}
\setlength{\unitlength}{1cm}
\begin{picture}(11,7)(2.5,0)
\put(0,0){\includegraphics[width=0.5\textwidth]{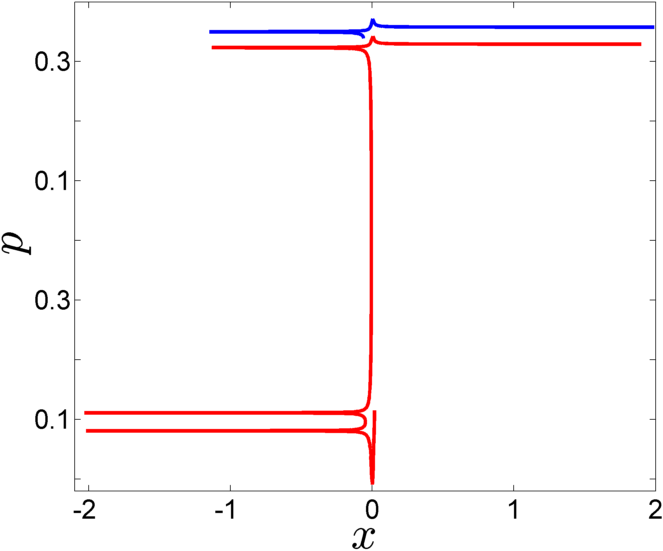}}
\put(0.9,2.3){\includegraphics[width=0.21\textwidth]{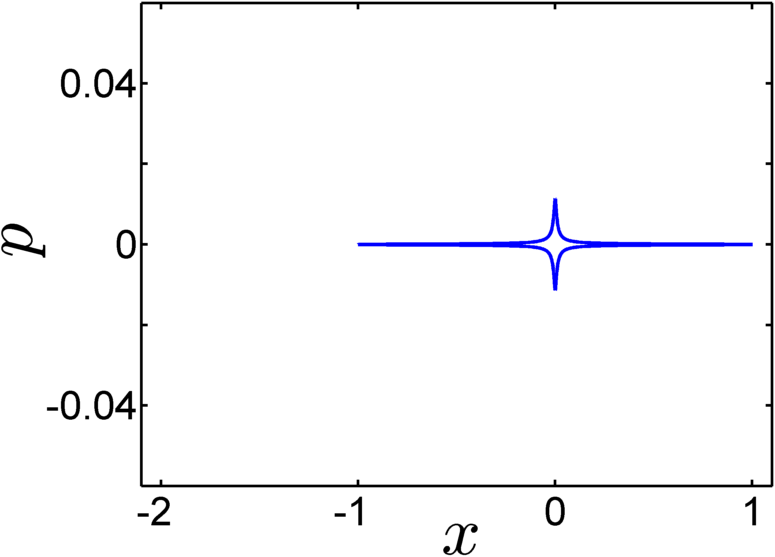}}
\put(2.0,4.3){$\mathcal{O}$}
\put(4.7,2.3){\includegraphics[width=0.21\textwidth]{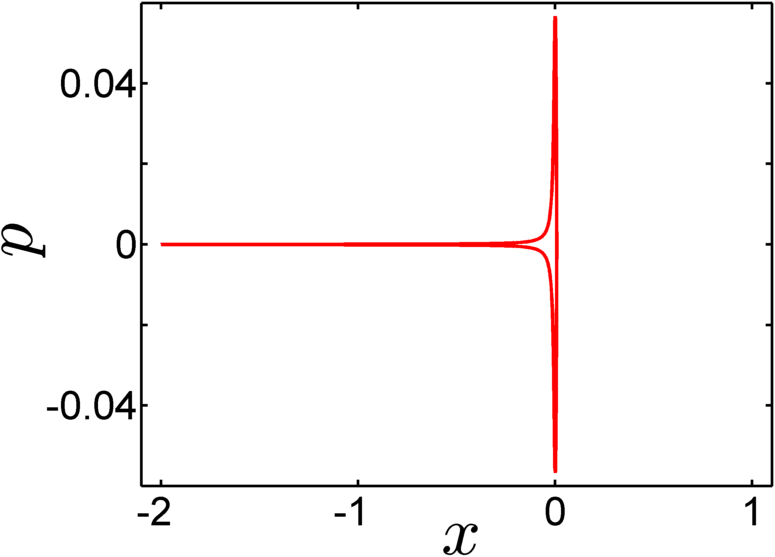}}
\put(5.7,4.3){$\mathcal{O}_-$}
\put(8.7,0.2){\includegraphics[width=0.45\textwidth]{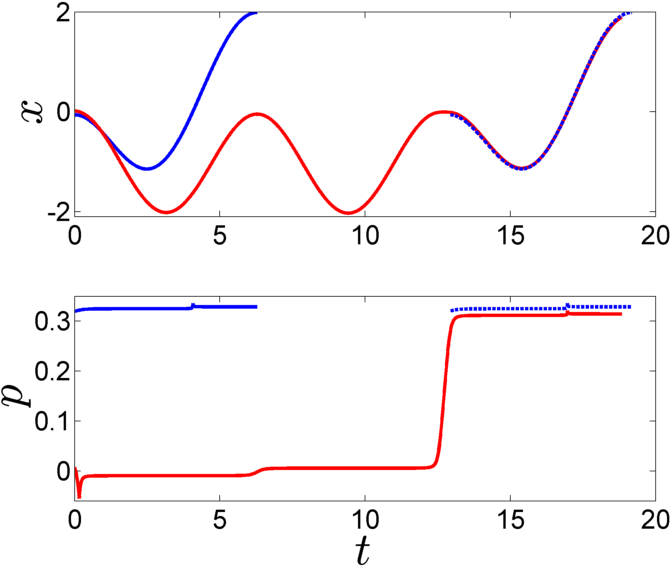}}
\end{picture}
\caption{(color online) Left: Two typical $\Delta \varepsilon$-maximizing trajectories in $x$-$p$ phase space. An immediate recollision is in black (blue), and a third-cycle delayed recollision is in grey (red). Left inset: Periodic orbit $\mathcal{O}$. Right inset: Periodic orbit $\mathcal{O}_-$. Right:  $\Delta \varepsilon$-maximizing trajectories of the left figure plotted as $x$ vs.\ $t$, upper, and $p$ vs.\ $t$, lower. The immediate recollision is also shown as a dotted black (blue) line, translated in time so its laser phase matches that of the delayed recollision. $x$ is in units of $E_0/\omega^2$, $p$ is in units of $E_0/\omega$, $t$ is in radians.}\label{recollisions}
\end{figure}

The qualitative character of both immediate and delayed recolliding trajectories can be understood by considering some periodic orbits of the system.
The most relevant periodic orbits are $\mathcal{O}$ and $\mathcal{O}_-$, shown as the insets on the left plot of Fig.~\ref{recollisions}.
The importance of $\mathcal{O}$ for recollisions was first described in Ref.~\cite{Kamo14}, and in Ref.~\cite{Norm15} it is shown that $\mathcal{O}_-$ (and $\mathcal{O}_+$, its symmetric copy via 
 symmetry \eqref{symmetry}) underlie ionization stabilization in ultra-intense laser fields.
Due to our choice of Poincar\'e section and our interest in initial conditions beginning near the core, we need not consider $\mathcal{O}_+$ here.
Each orbit has a period of exactly one laser cycle.
For the parameters we have chosen, both $\mathcal{O}$ and $\mathcal{O}_-$ are hyperbolic, i.e.\ they are unstable.
Thus, initial conditions in the vicinity of these orbits may take trajectories qualitatively similar to the orbits for some time, before eventually ionizing.

This is precisely what we observe when we examine both the immediate and delayed recollision trajectories in phase space and compare them with the periodic orbits.
In the left panel of Fig.~\ref{recollisions}, we have plotted an energetic immediate recollision and an energetic third-cycle delayed recollision.
The immediate recollision resembles the motion of $\mathcal{O}$, as was noted in Ref.\cite{Kamo14}.
$\mathcal{O}$'s main characteristics are motion between $x \approx-E_0/\omega^2$ and $x \approx E_0/\omega^2$, with a small momentum kick due to the Coulomb field upon recollision.
Correspondingly, the energetic immediate recollision also travels out to $x \approx -E_0/\omega^2$, before recolliding, receiving a small momentum kick, and ionizing forever.
Delayed recollisions on the other hand spend some time following the orbit $\mathcal{O}_-$, before reaching the part of phase space where immediate recollisions live and the motion of $\mathcal{O}$ dominates.
$\mathcal{O}_-$ is distinguished by its motion between $x \approx -2 E_0/\omega^2$ and $x \approx 0$, with a comparatively large momentum kick at the core.
Inspection of the third-cycle delayed recollision confirms that it has exactly this behavior for two laser cycles.
After the end of the second laser cycle, the momentum kick received by the electron is so large that the trajectory is moved into the region of phase space from which the energetic immediate recollisions originate.
Thus, delayed recollisions follow the motion of $\mathcal{O}_-$ for some number of laser cycles until they are converted into immediate recollisions.
At this stage, their motion is the same as an immediate recollision.
This can be observed from the right panels of Fig.~\ref{recollisions}, where we have plotted the energetic recollisions of the left panel as both $x$ vs.\ $t$ and $p$ vs.\ $t$, with the immediate recollision also translated forward in time so that its laser phase matches that of the delayed recollision.
Clearly, these trajectories nearly overlap, illustrating that the delayed recollisions eventually become immediate recollisions.

The arrangement of the delayed recollision initial conditions in phase space and their transport to the realm of immediate recollisions is determined by the stable and unstable manifolds, respectively, of $\mathcal{O}$.
The stable manifold $\mathcal{W}^s$ of a periodic orbit $\mathcal{O}$ is defined as the set of initial conditions that approach $\mathcal{O}$ as $t \rightarrow \infty$. Conversely, the unstable manifold $\mathcal{W}^u$ is the set of initial conditions that approach $\mathcal{O}$ as $t \rightarrow -\infty$, meaning those trajectories escape from the orbit moving forward in time.
It was already shown in Ref.~\cite{Kamo14} that $\mathcal{O}$'s invariant manifolds regulate the recollision dynamics.
There, the initial conditions leading to many recollisions were found to be concentrated near $\mathcal{W}^s$, and trajectories would follow $\mathcal{W}^u$ on their way to ionization.
It turns out that these manifolds also organize the delayed recollisions: delayed recollisions begin near $\mathcal{O}$'s stable manifold and then follow its unstable manifold to eventually recollide with the core.
The former is seen in the right panel of Fig.~\ref{poincare iterations}, where we show $\mathcal{W}^s$~\footnote{We have not marked the location of $\mathcal{O}$ on the figure because on this Poincar\'e section $\mathcal{O}$ is located at approximately $(x,p) = (E_0/\omega^2,0)$, far from the core.}.
$\mathcal{W}^s$ is obtained from $\mathcal{W}^u$, which we computed using the algorithm of Ref.~\cite{Hobs93}, by the time-reversal symmetry $(x,p,\phi) \rightarrow (x,-p,\pi - \phi)$.
Comparing the left and right upper panels of Fig.~\ref{poincare iterations} shows that all the delayed recollisions come from a region of phase space in which the curves of the stable manifold are dense (compared with other regions of phase space).
Looking at the delayed recollision region more closely in the lower panel of Fig.~\ref{poincare iterations} shows that the stable manifold actually separates regions with differing recollision delay times.
This is seen at the boundaries between immediate/second cycle recollisions, second/third cycle recollisions, and third/fourth cycle recollisions.
We expect that if the manifold is computed for longer times then it would also be seen to separate the higher order recollisions.
We note that there are parts of the manifold that are entirely contained in a region of a particular recollision delay time.
While these parts of the manifold are not relevant for separating initial conditions with different delay times, we expect they separate trajectories with another type of qualitative difference; for example, they may separate regions of initial conditions that have differing total numbers of recollisions, as in Ref.~\cite{Kamo14}.
Additionally, the delayed recollisions are indeed located near the orbit $\mathcal{O}_-$, as was suggested earlier by noting the similarities of the delayed recollision trajectories to the periodic orbit.

\begin{figure}
\setlength{\unitlength}{1cm}
\begin{picture}(11,10)
\put(0,0){\includegraphics[width=0.7\textwidth]{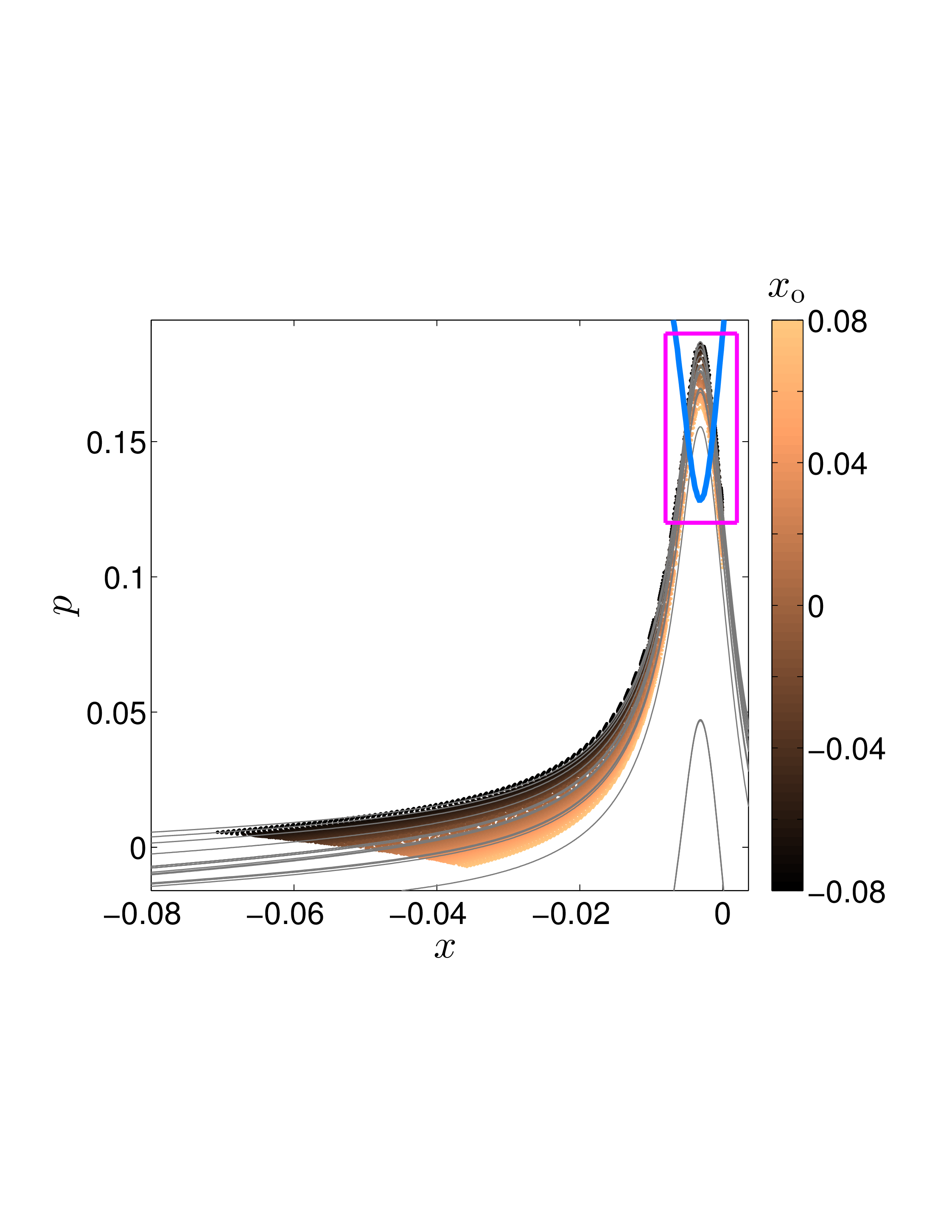}}
\put(1.6,4.5){\includegraphics[width=0.38\textwidth]{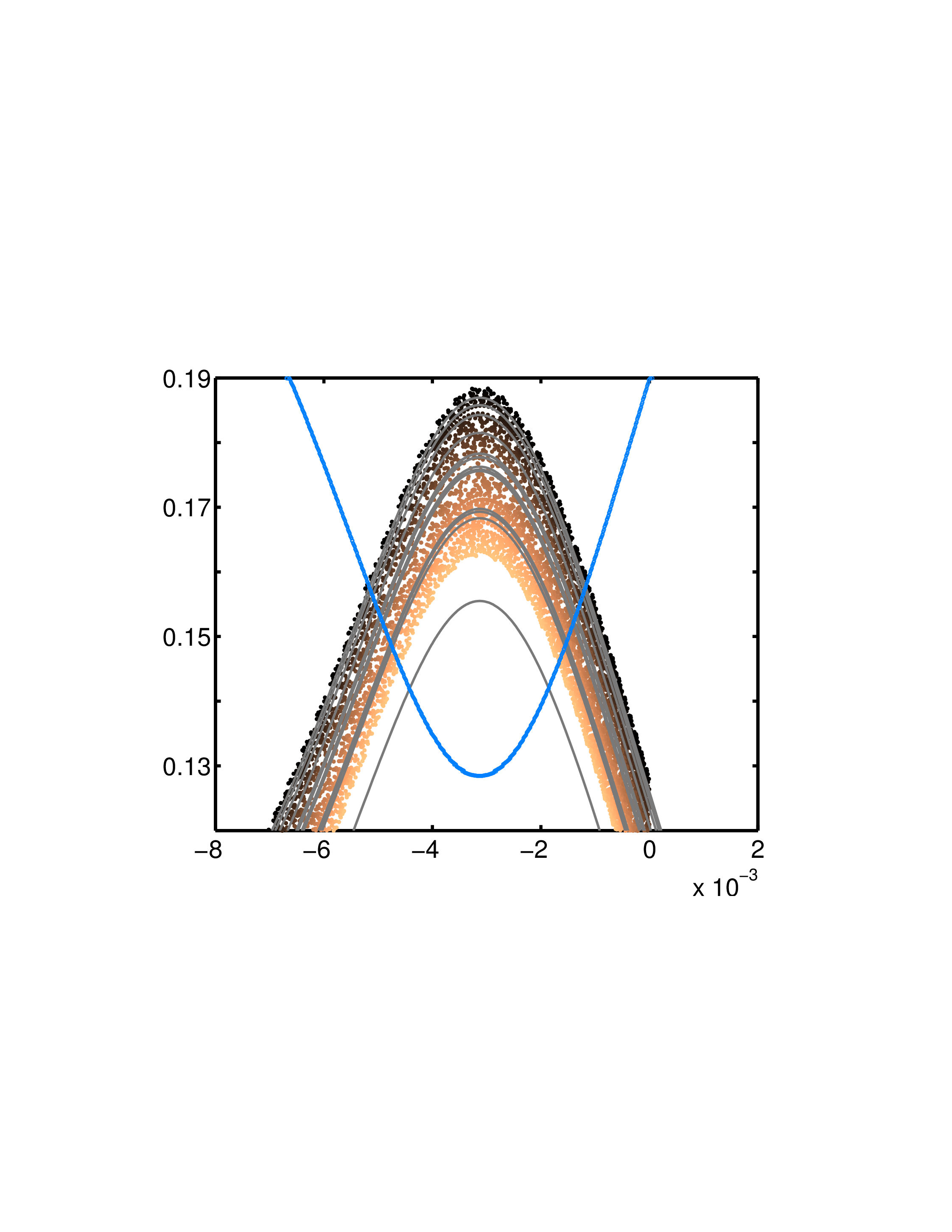}}
\end{picture}
\vspace{-90pt}
\caption{(color online) $\mathcal{P}(x_0,p_0)$ for the initial conditions $(x_0,p_0)$ in Fig.\ref{poincare iterations} leading to a second-cycle delayed recollision. Color scale shows $x_0$. Thick grey (light blue) curve is the initial conditions leading to a maximum $\Delta \varepsilon$ immediate recollision. Thin grey curve is $\mathcal{W}^u$ of $\mathcal{O}$ . Inset: Zoom of the grey (magenta)-framed region. $x$ is in units of $E_0/\omega^2$, $p$ is in units of $E_0/\omega$.}\label{fig:2ndcycle}
\end{figure}

When the initial conditions leading to delayed recollision are integrated, they follow $\mathcal{W}^u$, the unstable manifold of $\mathcal{O}$, and this process underlies our observations regarding the maximum energy available to a delayed recollision and the particular dependence of the maximum energy on $x_0$.
Consider an initial condition leading to an $n$-cycle delayed recollision.
By definition, after one iteration of the Poincar\'e map, it must move into the region of phase space from which $(n - 1)$-cycle recollisions originate, with $x < 0$.
This will happen repeatedly until the iteration immediately prior to recollision, when the trajectory reaches the immediate recollision region of phase space.
Additionally, the trajectories accomplish this motion by following $\mathcal{W}^u$.
Therefore, the recollision energies accessible to delayed recollisions are the immediate recollision energies in the vicinity of the unstable manifold.
These can be read off by looking at the upper right panel of Fig.~\ref{poincare energies}.
$\mathcal{W}^u$ enters the immediate recollision region above the region where $\mathcal{W}^s$ is dense and delayed recollisions proliferate.
We observe that in this region $\mathcal{W}^u$ is thickest for smaller $p_0$, and in this region the recollision energies are low, $\sim 0.5~U_p$.
This explains why most of the delayed recollisions recollide with an energy $\lesssim 0.5~U_p$, as we remarked earlier in reference to the lower left panel of Fig.~\ref{poincare energies}.
However, $\mathcal{W}^u$ does enter the region of high energy immediate recollisions, and in fact it is seen to intersect the curve of maximum $\Delta \varepsilon$ immediate recollisions in the lower right panel of Fig.~\ref{poincare energies}.
The delayed recollisions that end up in the region near this intersection are the delayed recollisions maximizing $\Delta \varepsilon$.

We look more carefully at the intersection of $\mathcal{W}^u$ with the curve of $\Delta \varepsilon$-maximizing immediate recollisions in Fig.~\ref{fig:2ndcycle}.
Here we have also plotted many second-cycle recollisions on their first return to Poincar\'e section \eqref{poincare condition}.
Firstly, the points are seen clearly to lie very close to $\mathcal{W}^u$, confirming our earlier claim that the delayed recollisions follow $\mathcal{W}^u$.
Next, we observe that there are always continuous gradients of $x_0$ perpendicular to the manifold, including in the region where the manifold intersects the curve of maximum energy immediate recollisions.
This region is magnified in the inset of Fig.~\ref{fig:2ndcycle}.
This is exactly why for any $x_0$, it is possible to find a second-cycle delayed recollision with $\Delta \varepsilon$ exceeding $3.17~U_p$. 
The energy of these delayed recollisions is
\begin{equation}\label{dE delayed}
\Delta \varepsilon_\text{max} = \Delta \varepsilon_\text{max,imm} (x_p) + \Delta V (x_p,x_0),
\end{equation}
where $\Delta \varepsilon_\text{max,imm} (x)$ is the maximum energy of an immediate recollision beginning at $x$, $x_p$ is the position of the delayed recollision's return to the Poincar\'e section on the cycle immediately prior to its recollision, and $\Delta V (x_p,x_0)$ is the change in energy between the delayed recollision at $t=0$ and the time it returned to the Poincar\'e section prior to recollision.
Because the section condition is zero velocity, there is no change in kinetic energy so $\Delta V$ is just the change in Coulomb potential energy.
Therefore we can obtain an estimate of the minimal $\Delta \varepsilon_\text{max}$ for a delayed recollision by taking $\Delta V (x_p,x_0) = -\epsilon/a$ and $\Delta \varepsilon_\text{max,imm}$ to be the minimal maximum energy accessible to an immediate recollision in the vicinity of $\mathcal{W}^u$.
This estimate is plotted for second-cycle delayed recollisions in Fig.~\ref{fig:dEvsx0}, and does a good job of providing a ``cutoff" for the minimal maximum energy second-cycle delayed recollision.
The remaining shape of the maximum energies for the delayed recollisions resembles the Coulomb potential and is mostly due to $\Delta V(x_p,x_0)$.
Intuitively, because $x_p$ takes a quite limited range of values, $\Delta \varepsilon_\text{max,imm}(x_p)$ is approximately constant and $\Delta V(x_p,x_0) \approx V(\tilde{x}_p) - V(x_0)$, where $\tilde{x}_p$ is some position where $\mathcal{W}^u$ intersects the maximum energy immediate recollision curve.
This explains why the delayed recollision energy curve approximately takes the shape of minus the Coulomb potential.

\begin{figure}
\includegraphics[width=0.55\textwidth]{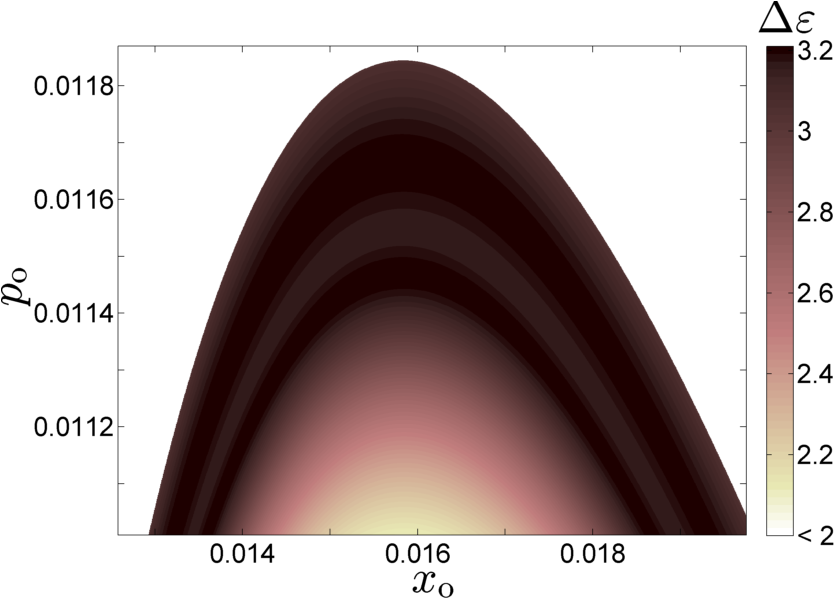}
\caption{(color online) Energy exchange $\Delta \varepsilon$ at first recollision as a function of initial conditions $(x_0,p_0)$. A region of second-cycle delayed recollisions is highly magnified here, revealing two local maxima in $\Delta \varepsilon$. Second-cycle delayed recollision initial conditions are in the colored region, immediate recollisions in the white region. Darker colors correspond to higher recollision energies. $x_0$ is in units of $E_0/\omega^2$, $p_0$ is in units of $E_0/\omega$, $\Delta \varepsilon$ is in units of $U_p$.}\label{fig:2ridge}
\end{figure}

In Fig.~\ref{fig:2ndcycle}, $\mathcal{W}^u$ is seen to intersect the curve of $\Delta \varepsilon$-maximizing immediate recollisions in two places.
In between the two intersections, $\mathcal{W}^u$ crosses into a region of lower energy recollisions, seen in the lower right panel of Fig.~\ref{poincare energies}, and thus we should expect that there should actually be two energy maximizing delayed recollisions for each $x_0$.
Indeed this is what we see, and this is the cause for the thickness of the set of points giving energy maximizing delayed recollisions in Fig.~\ref{fig:dEvsx0}.
This is seen even more clearly in Fig.~\ref{fig:2ridge}, where we have magnified a region containing second-cycle delayed recollision initial conditions and shown their energy exchange upon their first recollisions.
We observe two ridges, corresponding to the predicted two maximum energy recollisions for each $x_0$.

It is also possible to find higher order delayed recollisions of an arbitrary delay that return with an energy at least at the delayed recollision cutoff.
As we have argued, every laser cycle a delayed $n$-cycle recollision will move into the region of phase space from which the $n-1$-cycle recollisions originate, with $x < 0$.
Therefore, eventually any $n$-cycle delayed recollision will end up in the second-cycle delayed recollision region of phase space with $x < 0$, i.e.\ two laser cycles before its recollision.
But we have just shown that for any $x_0$, there are second cycle delayed recollisions that recollide with an energy above the delayed recollision cutoff.
Thus, every laser cycle, there are delayed recollisions arriving that bring an energy near $3.17~U_p$ to the core.
Though the SFA does a reasonable job of providing the energy cutoff, these kinds of trajectories are absent in the SFA, and are a direct consequence of the ion's Coulomb field and the organization of the dynamics by the periodic orbits $\mathcal{O}$ and $\mathcal{O}_\pm$ and the invariant manifolds of $\mathcal{O}$.

\section*{Conclusion}
In summary, we have shown that even in the high laser intensity regime where the strong field approximation is expected to hold, the Coulomb field significantly impacts the dynamics by allowing delayed recollisions to bring high energy to the core region. We reported the effectiveness of Coulomb focusing in a one-dimensional model, which is usually attributed to higher dimensional models.
We found that while the SFA gives adequate estimates of the maximum possible return energy in the strong field regime, it misses the behavior of the electron for times beyond the first laser cycle, when the Coulomb field causes trajectories to continue to recollide with energies near $3.17~U_p$.
We have unraveled the dynamical organization of these delayed recollisions by looking at specific periodic orbits and their invariant manifolds. 
The delayed recollision trajectories that we focused on are important not only because of the high energy they may recollide with, but also because they spend potentially many laser cycles near the core.
Thus, they have many opportunities to exchange energy with the ion and the electromagnetic field (due to the acceleration imparted on such electrons by the Coulomb force).
We expect that the delayed recollisions are the mechanism underlying HHG and thus explain the observation of the appearance of the plateau and $\sim3.17~U_p$ high-harmonic cutoff only $\emph{after}$ the first laser cycle, as in Ref.~\cite{Kamo14}.

\begin{acknowledgments}

The research leading to these results has received funding from the People Program (Marie Curie Actions) of the European Union’s Seventh Framework Program No. FP7/2007-2013/ under REA Grant No. 294974.
S.A.B.\ and T.U.\ acknowledge funding from the NSF (Grant No. 1304741).
S.A.B. acknowledges Burak Budanur for helpful discussions.

\end{acknowledgments}

% \bibliography{bibliography.bib}

\begin{thebibliography}{21}%
\makeatletter
\providecommand \@ifxundefined [1]{%
 \@ifx{#1\undefined}
}%
\providecommand \@ifnum [1]{%
 \ifnum #1\expandafter \@firstoftwo
 \else \expandafter \@secondoftwo
 \fi
}%
\providecommand \@ifx [1]{%
 \ifx #1\expandafter \@firstoftwo
 \else \expandafter \@secondoftwo
 \fi
}%
\providecommand \natexlab [1]{#1}%
\providecommand \enquote  [1]{``#1''}%
\providecommand \bibnamefont  [1]{#1}%
\providecommand \bibfnamefont [1]{#1}%
\providecommand \citenamefont [1]{#1}%
\providecommand \href@noop [0]{\@secondoftwo}%
\providecommand \href [0]{\begingroup \@sanitize@url \@href}%
\providecommand \@href[1]{\@@startlink{#1}\@@href}%
\providecommand \@@href[1]{\endgroup#1\@@endlink}%
\providecommand \@sanitize@url [0]{\catcode `\\12\catcode `\$12\catcode
  `\&12\catcode `\#12\catcode `\^12\catcode `\_12\catcode `\%12\relax}%
\providecommand \@@startlink[1]{}%
\providecommand \@@endlink[0]{}%
\providecommand \url  [0]{\begingroup\@sanitize@url \@url }%
\providecommand \@url [1]{\endgroup\@href {#1}{\urlprefix }}%
\providecommand \urlprefix  [0]{URL }%
\providecommand \Eprint [0]{\href }%
\providecommand \doibase [0]{http://dx.doi.org/}%
\providecommand \selectlanguage [0]{\@gobble}%
\providecommand \bibinfo  [0]{\@secondoftwo}%
\providecommand \bibfield  [0]{\@secondoftwo}%
\providecommand \translation [1]{[#1]}%
\providecommand \BibitemOpen [0]{}%
\providecommand \bibitemStop [0]{}%
\providecommand \bibitemNoStop [0]{.\EOS\space}%
\providecommand \EOS [0]{\spacefactor3000\relax}%
\providecommand \BibitemShut  [1]{\csname bibitem#1\endcsname}%
\let\auto@bib@innerbib\@empty
%</preamble>
\bibitem [{\citenamefont {Kuchiev}(1987)}]{Kuch87}%
  \BibitemOpen
  \bibfield  {author} {\bibinfo {author} {\bibfnamefont {M.~Y.}\ \bibnamefont
  {Kuchiev}},\ }\href@noop {} {\bibfield  {journal} {\bibinfo  {journal}
  {JETP~Lett.}\ }\textbf {\bibinfo {volume} {45}},\ \bibinfo {pages} {404}
  (\bibinfo {year} {1987})}\BibitemShut {NoStop}%
\bibitem [{\citenamefont {Corkum}(1993)}]{Cork93}%
  \BibitemOpen
  \bibfield  {author} {\bibinfo {author} {\bibfnamefont {P.~B.}\ \bibnamefont
  {Corkum}},\ }\href@noop {} {\bibfield  {journal} {\bibinfo  {journal}
  {Phys.~Rev.~Lett.}\ }\textbf {\bibinfo {volume} {71}},\ \bibinfo {pages}
  {1994} (\bibinfo {year} {1993})}\BibitemShut {NoStop}%
\bibitem [{\citenamefont {Schafer}\ \emph {et~al.}(1993)\citenamefont
  {Schafer}, \citenamefont {Yang}, \citenamefont {DiMauro},\ and\ \citenamefont
  {Kulander}}]{Scha93}%
  \BibitemOpen
  \bibfield  {author} {\bibinfo {author} {\bibfnamefont {K.~J.}\ \bibnamefont
  {Schafer}}, \bibinfo {author} {\bibfnamefont {B.}~\bibnamefont {Yang}},
  \bibinfo {author} {\bibfnamefont {L.~F.}\ \bibnamefont {DiMauro}}, \ and\
  \bibinfo {author} {\bibfnamefont {K.~C.}\ \bibnamefont {Kulander}},\
  }\href@noop {} {\bibfield  {journal} {\bibinfo  {journal} {Phys.~Rev.~Lett.}\
  }\textbf {\bibinfo {volume} {70}},\ \bibinfo {pages} {1599} (\bibinfo {year}
  {1993})}\BibitemShut {NoStop}%
\bibitem [{\citenamefont {Becker}\ \emph {et~al.}(2012)\citenamefont {Becker},
  \citenamefont {Liu}, \citenamefont {Ho},\ and\ \citenamefont
  {Eberly}}]{Beck12}%
  \BibitemOpen
  \bibfield  {author} {\bibinfo {author} {\bibfnamefont {W.}~\bibnamefont
  {Becker}}, \bibinfo {author} {\bibfnamefont {X.}~\bibnamefont {Liu}},
  \bibinfo {author} {\bibfnamefont {P.~J.}\ \bibnamefont {Ho}}, \ and\ \bibinfo
  {author} {\bibfnamefont {J.~H.}\ \bibnamefont {Eberly}},\ }\href@noop {}
  {\bibfield  {journal} {\bibinfo  {journal} {Rev. Mod. Phys.}\ }\textbf
  {\bibinfo {volume} {84}},\ \bibinfo {pages} {1011} (\bibinfo {year}
  {2012})}\BibitemShut {NoStop}%
\bibitem [{\citenamefont {Lewenstein}\ \emph {et~al.}(1994)\citenamefont
  {Lewenstein}, \citenamefont {Balcou}, \citenamefont {Ivanov}, \citenamefont
  {L'Huillier},\ and\ \citenamefont {Corkum}}]{Lewe94}%
  \BibitemOpen
  \bibfield  {author} {\bibinfo {author} {\bibfnamefont {M.}~\bibnamefont
  {Lewenstein}}, \bibinfo {author} {\bibfnamefont {P.}~\bibnamefont {Balcou}},
  \bibinfo {author} {\bibfnamefont {M.~Y.}\ \bibnamefont {Ivanov}}, \bibinfo
  {author} {\bibfnamefont {A.}~\bibnamefont {L'Huillier}}, \ and\ \bibinfo
  {author} {\bibfnamefont {P.~B.}\ \bibnamefont {Corkum}},\ }\href@noop {}
  {\bibfield  {journal} {\bibinfo  {journal} {Phys.~Rev.~A}\ }\textbf {\bibinfo
  {volume} {49}},\ \bibinfo {pages} {2117} (\bibinfo {year}
  {1994})}\BibitemShut {NoStop}%
\bibitem [{\citenamefont {Brabec}\ \emph {et~al.}(1996)\citenamefont {Brabec},
  \citenamefont {Ivanov},\ and\ \citenamefont {Corkum}}]{Brab96}%
  \BibitemOpen
  \bibfield  {author} {\bibinfo {author} {\bibfnamefont {T.}~\bibnamefont
  {Brabec}}, \bibinfo {author} {\bibfnamefont {M.~Y.}\ \bibnamefont {Ivanov}},
  \ and\ \bibinfo {author} {\bibfnamefont {P.~B.}\ \bibnamefont {Corkum}},\
  }\href@noop {} {\bibfield  {journal} {\bibinfo  {journal} {Phys.~Rev.~A}\
  }\textbf {\bibinfo {volume} {54}},\ \bibinfo {pages} {R2551} (\bibinfo {year}
  {1996})}\BibitemShut {NoStop}%
\bibitem [{\citenamefont {Shafir}\ \emph {et~al.}(2012)\citenamefont {Shafir},
  \citenamefont {Fabre}, \citenamefont {Higuet}, \citenamefont {Soifer},
  \citenamefont {Dagan}, \citenamefont {Descamps}, \citenamefont {M\'evel},
  \citenamefont {Petit}, \citenamefont {W\"orner}, \citenamefont {Pons},
  \citenamefont {Dudovich},\ and\ \citenamefont {Mairesse}}]{Shaf12}%
  \BibitemOpen
  \bibfield  {author} {\bibinfo {author} {\bibfnamefont {D.}~\bibnamefont
  {Shafir}}, \bibinfo {author} {\bibfnamefont {B.}~\bibnamefont {Fabre}},
  \bibinfo {author} {\bibfnamefont {J.}~\bibnamefont {Higuet}}, \bibinfo
  {author} {\bibfnamefont {H.}~\bibnamefont {Soifer}}, \bibinfo {author}
  {\bibfnamefont {M.}~\bibnamefont {Dagan}}, \bibinfo {author} {\bibfnamefont
  {D.}~\bibnamefont {Descamps}}, \bibinfo {author} {\bibfnamefont
  {E.}~\bibnamefont {M\'evel}}, \bibinfo {author} {\bibfnamefont
  {S.}~\bibnamefont {Petit}}, \bibinfo {author} {\bibfnamefont {H.~J.}\
  \bibnamefont {W\"orner}}, \bibinfo {author} {\bibfnamefont {B.}~\bibnamefont
  {Pons}}, \bibinfo {author} {\bibfnamefont {N.}~\bibnamefont {Dudovich}}, \
  and\ \bibinfo {author} {\bibfnamefont {Y.}~\bibnamefont {Mairesse}},\
  }\href@noop {} {\bibfield  {journal} {\bibinfo  {journal} {Phys.~Rev.~Lett.}\
  }\textbf {\bibinfo {volume} {108}},\ \bibinfo {pages} {203001} (\bibinfo
  {year} {2012})}\BibitemShut {NoStop}%
\bibitem [{\citenamefont {Shafir}\ \emph {et~al.}(2013)\citenamefont {Shafir},
  \citenamefont {Soifer}, \citenamefont {Vozzi}, \citenamefont {Johnson},
  \citenamefont {Hartung}, \citenamefont {Dube}, \citenamefont {Villeneuve},
  \citenamefont {Corkum}, \citenamefont {Dudovich},\ and\ \citenamefont
  {Staudte}}]{Shaf13}%
  \BibitemOpen
  \bibfield  {author} {\bibinfo {author} {\bibfnamefont {D.}~\bibnamefont
  {Shafir}}, \bibinfo {author} {\bibfnamefont {H.}~\bibnamefont {Soifer}},
  \bibinfo {author} {\bibfnamefont {C.}~\bibnamefont {Vozzi}}, \bibinfo
  {author} {\bibfnamefont {A.~S.}\ \bibnamefont {Johnson}}, \bibinfo {author}
  {\bibfnamefont {A.}~\bibnamefont {Hartung}}, \bibinfo {author} {\bibfnamefont
  {Z.}~\bibnamefont {Dube}}, \bibinfo {author} {\bibfnamefont {D.~M.}\
  \bibnamefont {Villeneuve}}, \bibinfo {author} {\bibfnamefont {P.~B.}\
  \bibnamefont {Corkum}}, \bibinfo {author} {\bibfnamefont {N.}~\bibnamefont
  {Dudovich}}, \ and\ \bibinfo {author} {\bibfnamefont {A.}~\bibnamefont
  {Staudte}},\ }\href {\doibase 10.1103/PhysRevLett.111.023005} {\bibfield
  {journal} {\bibinfo  {journal} {Phys. Rev. Lett.}\ }\textbf {\bibinfo
  {volume} {111}},\ \bibinfo {pages} {023005} (\bibinfo {year}
  {2013})}\BibitemShut {NoStop}%
\bibitem [{\citenamefont {Comtois}\ \emph {et~al.}(2005)\citenamefont
  {Comtois}, \citenamefont {Zeidler}, \citenamefont {Pépin}, \citenamefont
  {Kieffer}, \citenamefont {Villeneuve},\ and\ \citenamefont
  {Corkum}}]{Comt05}%
  \BibitemOpen
  \bibfield  {author} {\bibinfo {author} {\bibfnamefont {D.}~\bibnamefont
  {Comtois}}, \bibinfo {author} {\bibfnamefont {D.}~\bibnamefont {Zeidler}},
  \bibinfo {author} {\bibfnamefont {H.}~\bibnamefont {Pépin}}, \bibinfo
  {author} {\bibfnamefont {J.~C.}\ \bibnamefont {Kieffer}}, \bibinfo {author}
  {\bibfnamefont {D.~M.}\ \bibnamefont {Villeneuve}}, \ and\ \bibinfo {author}
  {\bibfnamefont {P.~B.}\ \bibnamefont {Corkum}},\ }\href
  {http://stacks.iop.org/0953-4075/38/i=12/a=008} {\bibfield  {journal}
  {\bibinfo  {journal} {J.~Phys.~B}\ }\textbf {\bibinfo {volume} {38}},\
  \bibinfo {pages} {1923} (\bibinfo {year} {2005})}\BibitemShut {NoStop}%
\bibitem [{\citenamefont {de~Bohan}\ \emph {et~al.}(2002)\citenamefont
  {de~Bohan}, \citenamefont {Piraux}, \citenamefont {Ponce}, \citenamefont
  {Ta\"ieb}, \citenamefont {V\'eniard},\ and\ \citenamefont {Maquet}}]{Boha02}%
  \BibitemOpen
  \bibfield  {author} {\bibinfo {author} {\bibfnamefont {A.}~\bibnamefont
  {de~Bohan}}, \bibinfo {author} {\bibfnamefont {B.}~\bibnamefont {Piraux}},
  \bibinfo {author} {\bibfnamefont {L.}~\bibnamefont {Ponce}}, \bibinfo
  {author} {\bibfnamefont {R.}~\bibnamefont {Ta\"ieb}}, \bibinfo {author}
  {\bibfnamefont {V.}~\bibnamefont {V\'eniard}}, \ and\ \bibinfo {author}
  {\bibfnamefont {A.}~\bibnamefont {Maquet}},\ }\href {\doibase
  10.1103/PhysRevLett.89.113002} {\bibfield  {journal} {\bibinfo  {journal}
  {Phys.~Rev.~Lett.}\ }\textbf {\bibinfo {volume} {89}},\ \bibinfo {pages}
  {113002} (\bibinfo {year} {2002})}\BibitemShut {NoStop}%
\bibitem [{\citenamefont {van~de Sand}\ and\ \citenamefont
  {Rost}(1999)}]{Sand99}%
  \BibitemOpen
  \bibfield  {author} {\bibinfo {author} {\bibfnamefont {G.}~\bibnamefont
  {van~de Sand}}\ and\ \bibinfo {author} {\bibfnamefont {J.~M.}\ \bibnamefont
  {Rost}},\ }\href@noop {} {\bibfield  {journal} {\bibinfo  {journal}
  {Phys.~Rev.~Lett.}\ }\textbf {\bibinfo {volume} {83}},\ \bibinfo {pages}
  {524} (\bibinfo {year} {1999})}\BibitemShut {NoStop}%
\bibitem [{\citenamefont {Ivanov}\ \emph {et~al.}(2005)\citenamefont {Ivanov},
  \citenamefont {Spanner},\ and\ \citenamefont {Smirnova}}]{Ivan05}%
  \BibitemOpen
  \bibfield  {author} {\bibinfo {author} {\bibfnamefont {M.~Y.}\ \bibnamefont
  {Ivanov}}, \bibinfo {author} {\bibfnamefont {M.}~\bibnamefont {Spanner}}, \
  and\ \bibinfo {author} {\bibfnamefont {O.}~\bibnamefont {Smirnova}},\
  }\href@noop {} {\bibfield  {journal} {\bibinfo  {journal} {J.~Mod.~Opt.}\
  }\textbf {\bibinfo {volume} {52}},\ \bibinfo {pages} {165} (\bibinfo {year}
  {2005})}\BibitemShut {NoStop}%
\bibitem [{\citenamefont {Bandrauk}\ \emph {et~al.}(2005)\citenamefont
  {Bandrauk}, \citenamefont {Chelkowski},\ and\ \citenamefont
  {Goudreau}}]{Band05}%
  \BibitemOpen
  \bibfield  {author} {\bibinfo {author} {\bibfnamefont {A.~D.}\ \bibnamefont
  {Bandrauk}}, \bibinfo {author} {\bibfnamefont {S.}~\bibnamefont
  {Chelkowski}}, \ and\ \bibinfo {author} {\bibfnamefont {S.}~\bibnamefont
  {Goudreau}},\ }\href@noop {} {\bibfield  {journal} {\bibinfo  {journal}
  {J.~Mod.~Opt.}\ }\textbf {\bibinfo {volume} {52}},\ \bibinfo {pages} {411 }
  (\bibinfo {year} {2005})}\BibitemShut {NoStop}%
\bibitem [{\citenamefont {Javanainen}\ \emph {et~al.}(1988)\citenamefont
  {Javanainen}, \citenamefont {Eberly},\ and\ \citenamefont {Su}}]{Java88}%
  \BibitemOpen
  \bibfield  {author} {\bibinfo {author} {\bibfnamefont {J.}~\bibnamefont
  {Javanainen}}, \bibinfo {author} {\bibfnamefont {J.~H.}\ \bibnamefont
  {Eberly}}, \ and\ \bibinfo {author} {\bibfnamefont {Q.}~\bibnamefont {Su}},\
  }\href@noop {} {\bibfield  {journal} {\bibinfo  {journal} {Phys.~Rev.~A}\
  }\textbf {\bibinfo {volume} {38}},\ \bibinfo {pages} {3430} (\bibinfo {year}
  {1988})}\BibitemShut {NoStop}%
\bibitem [{\citenamefont {Mauger}\ \emph {et~al.}(2009)\citenamefont {Mauger},
  \citenamefont {Chandre},\ and\ \citenamefont {Uzer}}]{Maug09_1}%
  \BibitemOpen
  \bibfield  {author} {\bibinfo {author} {\bibfnamefont {F.}~\bibnamefont
  {Mauger}}, \bibinfo {author} {\bibfnamefont {C.}~\bibnamefont {Chandre}}, \
  and\ \bibinfo {author} {\bibfnamefont {T.}~\bibnamefont {Uzer}},\ }\href@noop
  {} {\bibfield  {journal} {\bibinfo  {journal} {Phys.~Rev.~Lett.}\ }\textbf
  {\bibinfo {volume} {102}},\ \bibinfo {pages} {173002} (\bibinfo {year}
  {2009})}\BibitemShut {NoStop}%
\bibitem [{\citenamefont {Ruth}\ \emph {et~al.}(1983)\citenamefont {Ruth} \emph
  {et~al.}}]{Ruth83}%
  \BibitemOpen
  \bibfield  {author} {\bibinfo {author} {\bibfnamefont {R.~D.}\ \bibnamefont
  {Ruth}} \emph {et~al.},\ }\href@noop {} {\bibfield  {journal} {\bibinfo
  {journal} {IEEE Trans. Nucl. Sci}\ }\textbf {\bibinfo {volume} {30}},\
  \bibinfo {pages} {2669} (\bibinfo {year} {1983})}\BibitemShut {NoStop}%
\bibitem [{\citenamefont {Kamor}\ \emph {et~al.}(2014)\citenamefont {Kamor},
  \citenamefont {Chandre}, \citenamefont {Uzer},\ and\ \citenamefont
  {Mauger}}]{Kamo14}%
  \BibitemOpen
  \bibfield  {author} {\bibinfo {author} {\bibfnamefont {A.}~\bibnamefont
  {Kamor}}, \bibinfo {author} {\bibfnamefont {C.}~\bibnamefont {Chandre}},
  \bibinfo {author} {\bibfnamefont {T.}~\bibnamefont {Uzer}}, \ and\ \bibinfo
  {author} {\bibfnamefont {F.}~\bibnamefont {Mauger}},\ }\href {\doibase
  10.1103/PhysRevLett.112.133003} {\bibfield  {journal} {\bibinfo  {journal}
  {Phys.~Rev.~Lett.}\ }\textbf {\bibinfo {volume} {112}},\ \bibinfo {pages}
  {133003} (\bibinfo {year} {2014})}\BibitemShut {NoStop}%
\bibitem [{Note1()}]{Note1}%
  \BibitemOpen
  \bibinfo {note} {We omitted the maximum energies of the higher order delayed
  recollisions because the shrinking size of the regions of phase space leading
  to recollisions with increasingly longer delays (see Fig.~\ref {poincare
  iterations}) makes it challenging to accurately compute the initial condition
  that recollides with the maximum energy for a particular delay.}\BibitemShut
  {Stop}%
\bibitem [{\citenamefont {Norman}\ \emph {et~al.}(2015)\citenamefont {Norman},
  \citenamefont {Chandre}, \citenamefont {Uzer},\ and\ \citenamefont
  {Wang}}]{Norm15}%
  \BibitemOpen
  \bibfield  {author} {\bibinfo {author} {\bibfnamefont {M.~J.}\ \bibnamefont
  {Norman}}, \bibinfo {author} {\bibfnamefont {C.}~\bibnamefont {Chandre}},
  \bibinfo {author} {\bibfnamefont {T.}~\bibnamefont {Uzer}}, \ and\ \bibinfo
  {author} {\bibfnamefont {P.}~\bibnamefont {Wang}},\ }\href {\doibase
  10.1103/PhysRevA.91.023406} {\bibfield  {journal} {\bibinfo  {journal} {Phys.
  Rev. A}\ }\textbf {\bibinfo {volume} {91}},\ \bibinfo {pages} {023406}
  (\bibinfo {year} {2015})}\BibitemShut {NoStop}%
\bibitem [{Note2()}]{Note2}%
  \BibitemOpen
  \bibinfo {note} {We have not marked the location of $\protect \mathcal {O}$
  on the figure because on this Poincar\'e section $\protect \mathcal {O}$ is
  located at approximately $(x,p) = (E_0/\omega ^2,0)$, far from the
  core.}\BibitemShut {Stop}%
\bibitem [{\citenamefont {Hobson}(1993)}]{Hobs93}%
  \BibitemOpen
  \bibfield  {author} {\bibinfo {author} {\bibfnamefont {D.}~\bibnamefont
  {Hobson}},\ }\href {\doibase http://dx.doi.org/10.1006/jcph.1993.1002}
  {\bibfield  {journal} {\bibinfo  {journal} {J.~Comput.~Phys.}\ }\textbf
  {\bibinfo {volume} {104}},\ \bibinfo {pages} {14 } (\bibinfo {year}
  {1993})}\BibitemShut {NoStop}%
\end{thebibliography}

%merlin.mbs apsrev4-1.bst 2010-07-25 4.21a (PWD, AO, DPC) hacked
%Control: key (0)
%Control: author (8) initials jnrlst
%Control: editor formatted (1) identically to author
%Control: production of article title (-1) disabled
%Control: page (0) single
%Control: year (1) truncated
%Control: production of eprint (0) enabled
%

\end{document}